\documentclass[%
 reprint,
superscriptaddress,
 amsmath,amssymb,
 aps,
]{revtex4-2}
\usepackage{xcolor}
\usepackage{graphicx}
\usepackage{dcolumn}
\usepackage{bm}
\usepackage{tikz}
\usepackage[utf8]{inputenc}
\usepackage[T1]{fontenc}
\usepackage{textcomp}

\begin{document}


\title{Antiferromagnetism and Tightly Bound Cooper Pairs Induced by Kinetic Frustration}

\author{Yixin Zhang}
\affiliation{Department of Physics and Astronomy, University of Tennessee, Knoxville, Tennessee 37996, USA}
\affiliation{School of Physics, Peking University, Beijing 100871, China}

\author{Cristian D. Batista}
\affiliation{Department of Physics and Astronomy, University of Tennessee, Knoxville, Tennessee 37996, USA}
\affiliation{Neutron Scattering Division, Oak Ridge National Laboratory, Oak Ridge, TN, USA}

\author{Yang Zhang}
\affiliation{Department of Physics and Astronomy, University of Tennessee, Knoxville, Tennessee 37996, USA}
\affiliation{Min H. Kao Department of Electrical Engineering and Computer Science, University of Tennessee, Knoxville, Tennessee 37996, USA}


\begin{abstract}
Antiferromagnetism and superconductivity are often viewed as competing orders in correlated electron systems. Here, we demonstrate that kinetic frustration in hole motion facilitates their coexistence within the square-lattice repulsive Hubbard model. Combining exact analytical solutions on tailored geometries with large-scale numerical simulations, we reveal a robust pairing mechanism: holes on opposite sublattices behave as if they carry opposite effective charges due to spin singlet formation from kinetic frustration. This emergent property suppresses phase separation and fosters a coherent $d$-wave superconducting channel embedded within a long-range antiferromagnetic background. Our findings establish a minimal yet broadly applicable framework for stabilizing strong-coupling superconductivity in doped Mott insulators.
\end{abstract}

\maketitle


\section{Introduction}

The interplay between magnetism and superconductivity is a defining characteristic of unconventional superconductors, spanning high-$T_c$ cuprates~\cite{Lee2006}, heavy-fermion systems~\cite{Thompson2003}, organic conductors~\cite{Kanoda2011} and iron-based superconductors~\cite{Stewart2011}. In cuprates, superconductivity emerges upon doping an antiferromagnetic Mott insulator, prompting decades of theoretical efforts to uncover the microscopic origin of the resulting $d$-wave pairing state~\cite{Keimer2015}. 
The square-lattice repulsive Hubbard model has served as a paradigmatic framework for addressing this problem, capturing both the Mott insulating behavior at half-filling and the emergence of pairing correlations upon doping~\cite{Dagotto1994,Lee2006,Keimer2015}.
Despite its conceptual simplicity, the microscopic origin of superconductivity in the doped Hubbard model remains controversial. In the weak-coupling regime, antiferromagnetic spin fluctuations have been proposed as a natural pairing mechanism favoring $d$-wave symmetry~\cite{Bickers1989,Monthoux1991}. Conversely, numerical studies indicate $d$-wave pairing near antiferromagnetic order in the strong-coupling limit~\cite{Scalapino2007}, although these simulations also reveal a proclivity for phase separation~\cite{Emery1990,Moreo1991}, destabilizing uniform superconductivity and obscuring a unified understanding of the pairing mechanism.

More broadly, a long-standing puzzle persists across families of lightly doped antiferromagnetic insulators. While materials such as LaTiO$_3$, Sr$_2$IrO$_4$, V$_2$O$_3$, NiS$_2$, and Sr$_2$VO$_4$ remain non-superconducting even at ultralow temperatures, others---including electron-doped cuprates~\cite{Kastner1998,Lee2006,armitage2010progress,Keimer2015}, heavy-fermion systems like CeRh(Co)In$_5$~\cite{Stock2008}, and certain organic conductors~\cite{lefebvre2000mott}---exhibit a robust coexistence of antiferromagnetism and superconductivity. This contrast eludes conventional explanations based solely on strong superexchange interactions or low-dimensionality, suggesting that magnetic fluctuations \emph{alone} are insufficient to account for doping-induced superconductivity.

Identifying a clear pairing mechanism from numerical studies of the doped Hubbard model is notoriously difficult due to the intricate competition between AFM ordering and kinetic energy of doped holes, which leads to multiple intertwined orders. In the strong-coupling regime, superconductivity often competes with robust antiferromagnetic correlations inherited from the Mott insulating state, resulting in a proliferation of nearly degenerate ground states, including stripes, charge density waves, and pair-density wave states~\cite{timm2000doping,Scalapino2007,Berg2009,zheng2017stripe,huang2017numerical,gong2021robust,xu2024coexistence,chen2025global}. As a result, numerical signatures of superconductivity—such as long-range pairing correlations or anomalous response functions—are often entangled with these other ordering tendencies, complicating efforts to disentangle the intrinsic pairing mechanism. Moreover, finite-size effects and the lack of direct access to dynamical pairing interactions further hinder a definitive interpretation of the results, leaving open the fundamental question of what drives pairing in this paradigmatic model.

\begin{figure}
\includegraphics[width=\columnwidth]{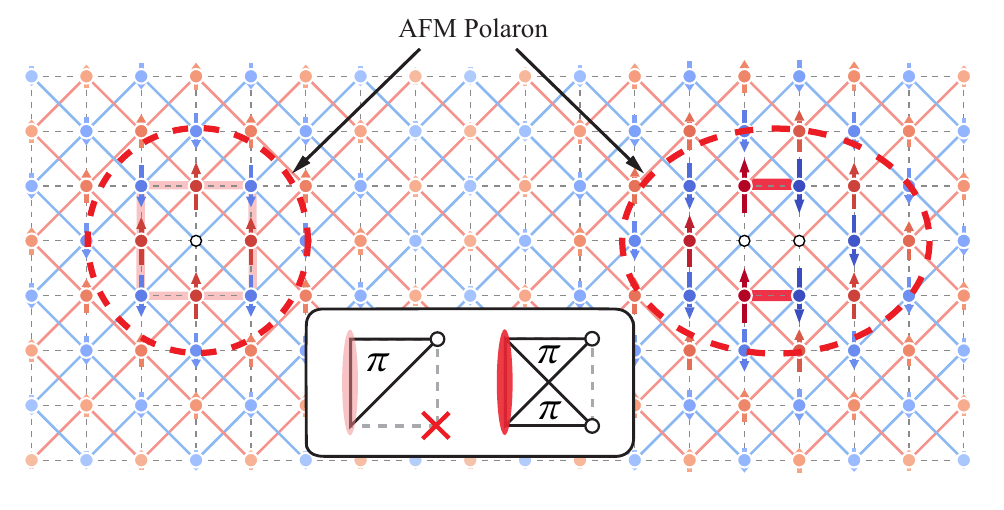}
\caption{\label{fig:1} Antiferromagnetic polaron induced by kinetic frustration. The two magnetic sublattices are shown with red and blue spins, polarized in opposite directions. Empty circles represent holes. As shown on the left, a strong singlet character---indicated by red bonds---is expected to form near each hole. This is because, as illustrated in the inset, a singlet state effectively introduces a $\pi$-flux (relative to the triplet state) in the triangle formed by the singlet and the hole. The presence of this $\pi$-flux lowers the kinetic energy of the hole by alleviating KF and generates an attractive potential for a second hole. This cooperative effect manifests as a pronounced singlet character along the two bonds parallel to the adjacent hole pair, as depicted in the inset and on the right side of the figure.
 }
\end{figure}

Kinetic frustration (KF) has recently emerged as a central mechanism for generating strong pairing from purely repulsive interactions in doped fully polarized Mott insulators~\cite{isaev2010superconductivity,ZhangShangshun2018,Nazaryan2024}. In the strongly correlated regime, frustration of hole motion gives rise to an effective antiferromagnetic interaction between neighboring spins, producing a ``counter-Nagaoka effect'' in the infinite-$U$ limit~\cite{Haerter2005,sposetti_classical_2014,ciorciaro2023,kim2023exact}. This emergent interaction originates from the formation of a local singlet on triangular plaquettes occupied by a hole, which suppresses KF by inducing an effective $\pi$-flux through the triangle (see inset of Fig.~\ref{fig:1}). In polarized spin backgrounds, this mechanism leads to the formation of tightly bound hole-magnon composites, or spin polarons~\cite{ZhangShangshun2018,zhang2023}. Remarkably, the same kinetic interference responsible for polaron formation can also mediate a strong pairing interaction between polarons, giving rise to a novel superconducting phase termed \emph{magnonic superconductivity} in regimes of sufficiently strong KF~\cite{ZhangShangshun2018,Nazaryan2024}.

Here, we leverage the guiding principle of KF to reveal a robust microscopic mechanism that significantly reduces the complexity inherent in the doped Hubbard model. Our central insight is that KF \emph{simultaneously stabilizes both antiferromagnetic correlations and superconductivity} in the strongly coupling regime, thereby eliminating the primary source of competition that has hindered a clear identification of the pairing mechanism. By combining exact analytical solutions on simplified geometries with large-scale numerical simulations (up to lattices of $16\times16$ sites) of an extended square-lattice $t$--$J$ model with correlated hoppings, we demonstrate that frustration of hole motion---arising from competing nearest-neighbor (NN) and next-nearest-neighbor (NNN) hopping amplitudes, $t_1$ and $t_2$---gives rise to a superconducting phase coexisting with long-range AFM order. While the resulting pairing mechanism shares similarities with \emph{magnonic superconductivity} in a ferromagnetic background, a crucial distinction lies in the fact that, in our case, the AFM background is dynamically stabilized by the hole kinetic energy. As shown in Fig.~\ref{fig:1}, the associated polaronic effect enhances the singlet character of spin pairs on bonds that form triangles with the doped hole to suppress the KF. When a second hole occupies the opposite sublattice, it experiences an effective attraction of order $t_1$ when located adjacent to the first hole, driven by the cooperative enhancement of singlet correlations on the two bonds parallel to the two holes. The resulting \emph{two-body hopping}  naturally generates strong pairing without requiring any bare attractive interaction.

A hallmark of the mechanism described above is that Cooper pair formation is accompanied by the development of strong singlet character on the bonds surrounding the charge carriers. At elevated doping levels, this process is expected to drive a quantum phase transition from antiferromagnetic (AFM) order to a quantum paramagnetic state characterized by a spin gap and short-range magnetic correlations, with correlation lengths on the order of the lattice spacing. Notably, these magnetic features \emph{emerge only within the superconducting phase}. 

This picture is consistent with observations of superconductivity accompanied by enhanced singlet formation in various families of unconventional superconductors, including the high-\(T_c\) cuprates~\cite{Kastner1998,Lee2006,Keimer2015}, CeRh(Co)In\(_5\)~\cite{Stock2008}, and FeTe\(_{1-x}\)Se\(_x\)~\cite{Leiner2014}. The enhanced singlet correlations near mobile charge carriers can also be directly probed in ultracold fermionic atoms in optical lattices through spin-resolved quantum gas microscopy~\cite{xu2025neutral}.

Our results offer a new perspective on the emergence of superconductivity from purely repulsive interactions in doped Mott insulators and establish a minimal theoretical framework with broad relevance to strongly correlated quantum materials.



\section{N\'eel order and pairing}
\label{N&P}

\begin{figure}
\includegraphics[width=\columnwidth]{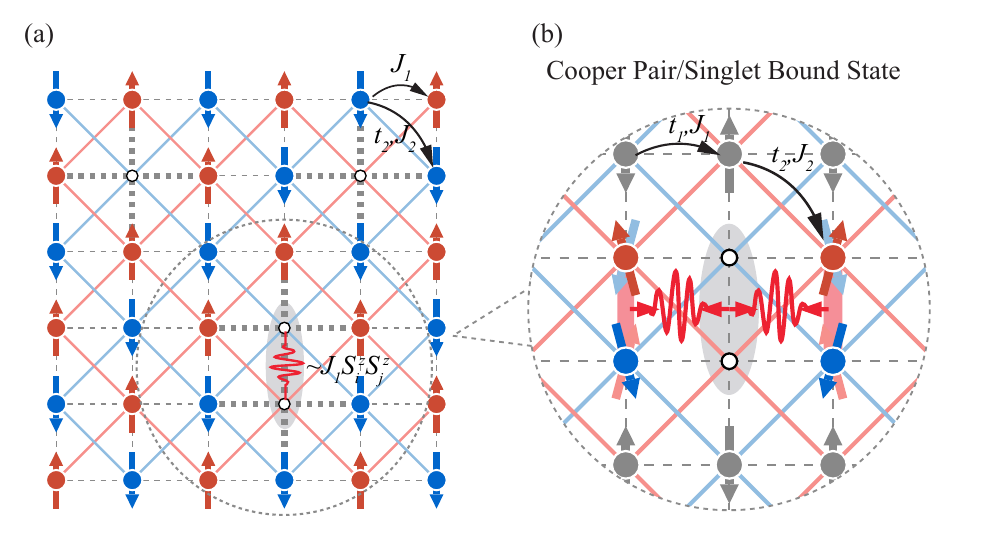}
\caption{\label{fig:2}  (a) Schematic diagram of the $t_2-J_{z}$ model's Néel ordered phase with two doped holes. Shaded ellipses indicate Cooper pairs of two holes, and red inward arrows on the wavy lines indicate attractive interactions. (b) Zoom-in figure illustrating the spin fluctuations around the hole pair. The shaded Cooper pair forms a bound state with singlet spin configuration connected by thick red shading in order lower its KE.}
\end{figure}

We begin by considering the large-$U$ limit of an extended Hubbard model on a square lattice including a NNN hopping amplitude $t_2$ required to frustrate the hole KE. In this limit ($U \gg |t_1|, |t_2|$), the low-energy physics of the Hubbard model is projected onto an effective generalized $t-J$ model $H = H_{t} + H_{J}+H_{\text{c-h}}$~\cite{chao1978canonical,trugman_interaction_1988, zhang1988effective}:

\begin{equation}
H_{t} = - \sum_{j l, \sigma} t_{jl} c_{j \sigma}^{\dagger} c_{l \sigma} ,
\label{eq:hamilt}
\end{equation}
\begin{equation}
H_{J} = \sum_{ j l} \frac{J_{jl}}{2} \left[S^z_j  S^z_l + \Delta (S^x_j  S^x_l + S^y_j  S^y_l) -\frac{n_j n_l}{4} \right],
\label{eq:hamilJ}
\end{equation}
\begin{eqnarray}
H_{\text{c-h}} &=& - \Delta\sum_{i, j, k, \sigma}^{i \neq k} \frac{t_{i j} t_{j k}}{U} [c_{i \sigma}^{\dagger} c_{k \sigma} n_{j \bar{\sigma}} - c_{i \sigma}^{\dagger} c_{k \bar{\sigma}} c_{j \bar{\sigma}}^{\dagger} c_{j \sigma}].
\label{eq:hamilcorr}
\end{eqnarray}
The NN and NNN hopping amplitudes are parameterized as $\lambda t_1$ and $t_2$, respectively, and they are assumed to be real. The parameter $\lambda$ is introduce to connect the case of interest ($\lambda = 1$ recovers the effective low-energy Hamiltonian of the Hubbard model) with an exactly solvable limit for $\lambda=0$. The corresponding NN and NNN exchange interactions are given by: $J_1 = 4 t_{1}^2 / U$ and $J_2 = 4 t_{2}^2 / U$. We also introduce an exchange anisotropy parameter $\Delta$, which reduces the spin exchange to the Ising type and simultaneously suppresses the correlated hopping term $H_{\text{c-h}}$, ensuring the full Hamiltonian $H$ smoothly interpolates between the isotropic low-energy effective model ($\Delta = 1$) and a pure $t-J_z$ limit ($\Delta = 0$). The spin degrees of freedom are represented by the $S=1/2$ spin operator $S^a_j \equiv \frac{1}{2} \sum_{\alpha \beta} c^{\dagger}_{j\alpha} \sigma^{a}_{\alpha \beta} c^{\;}_{j\beta}$ ($a=x,y,z$) and $n_j \equiv \sum_{\alpha}  c^{\dagger}_{j\alpha} c^{\;}_{j\alpha}$ denotes the charge density operator.

As shown in Fig.~\ref{fig:2}(a), the limit $\lambda = \Delta = 0$ retains only the $t_2$ hopping and the Ising exchange terms $J_{1} S^z_j S^z_l$ and $J_{2} S^z_j S^z_l$, rendering the two-hole ground state of $H$ exactly solvable. At infinite $U$, Nagaoka’s theorem~\cite{nagaoka_ferromagnetism_1966, tasaki_extension_1989, tasaki_nagaokas_1998} indicates that energy minimization requires the two holes to reside on distinct sublattices. The addition of weak Ising interactions $J_{1}$ and $J_{2}$ further stabilizes a bipartite antiferromagnetic order: ferromagnetic alignment within each sublattice and antiferromagnetic alignment between them, provided that $J_{2} < 0.5 J_{1}$~\cite{gangat_weak_2024}. In this regime, AFM order is jointly stabilized by the hole kinetic energy and the dominant Ising coupling $J_{1}$.

When two holes occupy nearest-neighbor sites on opposite sublattices, the number of disrupted antiferromagnetic bonds decreases from 8 to 7. This reduction generates an effective NN attraction of magnitude $J_{1}/2$. While this attraction suffices to form a two-hole bound state in two dimensions, the binding energy follows an exponential dependence as $e^{-4 \pi |t_2|/J_{1}}$ in the small $J_{1}/t_2$ limit. Consequently, the pair coherence length $\xi$ — characterizing the spatial extent of the bound state — vastly exceeds the lattice constant $a$ ($\xi \gg a$), indicating a weakly localized pair. As shown in subsequent sections, the inclusion of transverse spin fluctuations significantly amplifies the pairing strength, reducing $\xi$ to just a few lattice constants.




\section{Spin fluctuations}
To systematically quantify the role of spin fluctuations in pairing enhancement, we employ a two-stage approach to gradually introduce spin fluctuation into the classical Néel-ordered background. First, we activate nearest-neighbor hopping through the parameter $\lambda$, which scales the NN hopping amplitude. For $\lambda \ll 1$ {with $t_1^2 t_2 > 0$}, the counter-Nagaoka mechanism stabilizes antiparallel spin alignment for hole doping, as demonstrated in the supplementary material (SM). 

Remarkably, the approximate N\'eel order can be preserved even for moderate values up to $t_1 \sim 2 t_2$ in the $U\to \infty$ limit. At intermediate $t_2/t_1$ ratios, particularly around $t_2/t_1 \in (0.5, 0.7)$, both the kinetic hole motion and exchange interactions ($J_2<\frac{1}{2}J_1$~\cite{morita_quantum_2015}) cooperatively support N\'eel AFM order. For all subsequent calculations, we adopt $t_2 = 0.6 t_1$.

The second source of spin fluctuations arises from second-order processes: a finite value of $\Delta > 0$ activates both the spin exchange terms $(S^x_j S^x_l + S^y_j S^y_l)$ and the correlated hopping processes. Although these interactions introduce additional quantum fluctuations, they preserve the underlying antiferromagnetic Néel order in the undoped insulating phase, as confirmed by the established phase diagram of the square-lattice $J_1$--$J_2$ Heisenberg model~\cite{morita_quantum_2015}.

The central computational challenge arises from the exponential growth of the Hilbert space due to spin fluctuations. Motivated by the polaronic nature of hole motion illustrated in Fig.~\ref{fig:1}, and by the expectation that the polaron size remains small—since both exchange and kinetic energies favor antiferromagnetic ordering—we develop a constrained approach for hole dynamics. In this method, spin fluctuations are included only within restricted regions surrounding each hole~\cite{trugman_interaction_1988}, while maintaining a fixed total size of the fluctuation-allowed Hilbert space across different hole configurations and system sizes (see SM for details). This truncation strategy enables exact diagonalization calculations with Hilbert space dimensions that scale quadratically with system size, reaching up to $1.1 \times 10^9$ states for a $16 \times 16$ cluster.

To validate our truncated Hilbert space approach, we first benchmark it against the isotropic quantum limit ($\lambda = 1$ and $\Delta = 1$) with $U = 10 t_1$, which restores the full $SU(2)$ symmetry of the standard $t$--$J$--$t_{\text{c-h}}$ model. In particular, we compute the hole density-density correlation function:
\begin{equation}
    P_{i,j} = \langle\Psi_G | \bar{n}_{i} \bar{n}_{j} |\Psi_G\rangle, \label{hole-hole correlation function}
\end{equation}
where $\bar{n}_j \equiv 1 - n_j$ is the hole density operator and $\langle\Psi_G | \cdot |\Psi_G\rangle$ denotes the ground state expectation value. As an independent benchmark, we perform charge-conserving density matrix renormalization group (DMRG)~\cite{White1992} calculations using the same Hamiltonian on an 8-leg cylinder with $L_x = 16$ (see SM). The cylindrical geometry naturally avoids the complications associated with periodic boundary conditions (PBC) in DMRG and allows for system lengths sufficient to minimize edge-induced finite-size effects that can suppress hole density near open boundaries.

\begin{figure}
\includegraphics[width=\columnwidth]{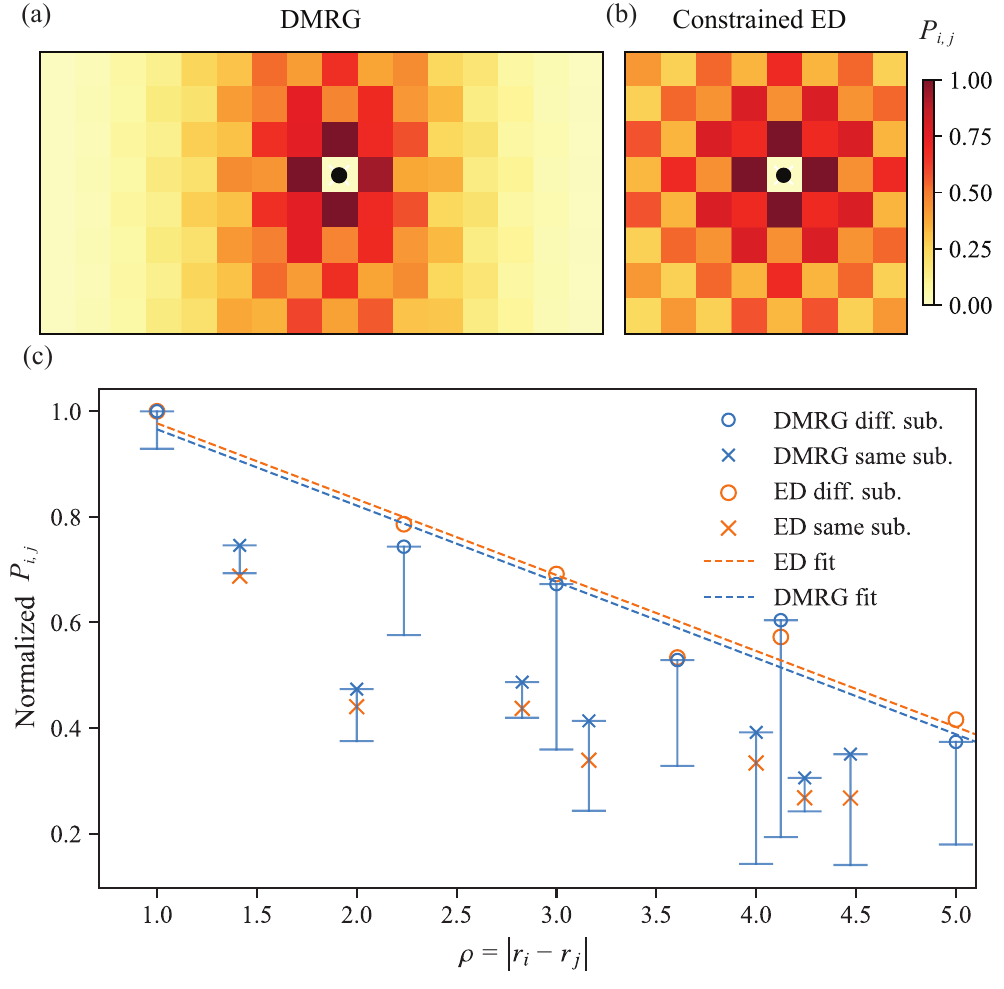}
\caption{\label{fig:3}(a) Hole-hole correlation $P_{i,j} = \langle\Psi_G | \bar{n}_{i} \bar{n}_{j} |\Psi_G\rangle$ calculated with ground state converged using U(1) DMRG on an 8-leg cylinder and keeping up to 8000 states per block. Position of site $i$: \(r_i\) is fixed at the black dot and position \(r_j\) varying across the lattice. We use $t_2/t_{1} = 0.6$, $\lambda = 1$, $\Delta = 1$, and $U = 10 t_{1}$ in both simulations. The correlation is calculated with one hole position fixed at the center, and normalized to $[0, 1]$. (b) Hole-hole correlation obtained through Hilbert space truncated ED calculation on an $8\times 8$ system with PB. (c) Comparison between ED and DMRG correlations as a function of distance. The error bars in DMRG represent the spread of correlations at the same distance due to boundary effects which lower hole density away from the center. Dashed lines are fitted with opposite sublattice correlations. 
}
\end{figure}

As shown in Fig.~\ref{fig:3}(c), we find quantitative agreement between ED and DMRG results. Both methods consistently reveal: (1) strong attractive interactions between holes; (2) robust $d_{x^2 - y^2}$-wave pairing correlations (see SM for details); and (3) striking sublattice-dependent correlation patterns. Notably, hole correlations are significantly enhanced when the holes occupy opposite sublattices compared to the same sublattice. This sublattice selectivity directly reflects the pairing mechanism of our exactly solvable parent model, where optimal hole binding occurs between distinct sublattices.

Beyond correlation functions, our constrained ED framework provides direct access to the two-hole binding energy - a critical metric for pairing strength. We define the binding gap as the energy difference between the ground state and first excited state within a fixed momentum sector: $\Delta_{g}(\bm{k}) = E_1(\bm{k}) - E_0(\bm{k})$. In the thermodynamic limit, the ground state energy $E_0(\bm{k})$ corresponds to a bound pair while the first excited state energy $E_1(\bm{k})$ corresponds to the lowest unpaired state in the scattering continuum. Thus, $\Delta_{g}(\bm{k})$ serves as a direct measure of the superconducting pairing gap.

To minimize finite-size effects, we implement twisted boundary conditions by choosing momentum $k = (0, 2\pi/L)$. This choice imposes distinct boundary conditions on the relative motion of the two-hole system: periodic along one spatial direction and antiperiodic along the orthogonal direction. Critically, this configuration suppresses spurious finite-size effects—specifically, the artificial binding gap inherent to non-interacting systems under conventional periodic PBC—thereby isolating interaction-driven pairing contributions (see SM for analytical justification). In the thermodynamic limit ($L \to \infty$), the $\bm{k} = (0, 2\pi/L)$ sector converges to $\bm{k} = 0$, where the Cooper pair ground state naturally resides.

\begin{figure}
\includegraphics[width=\columnwidth]{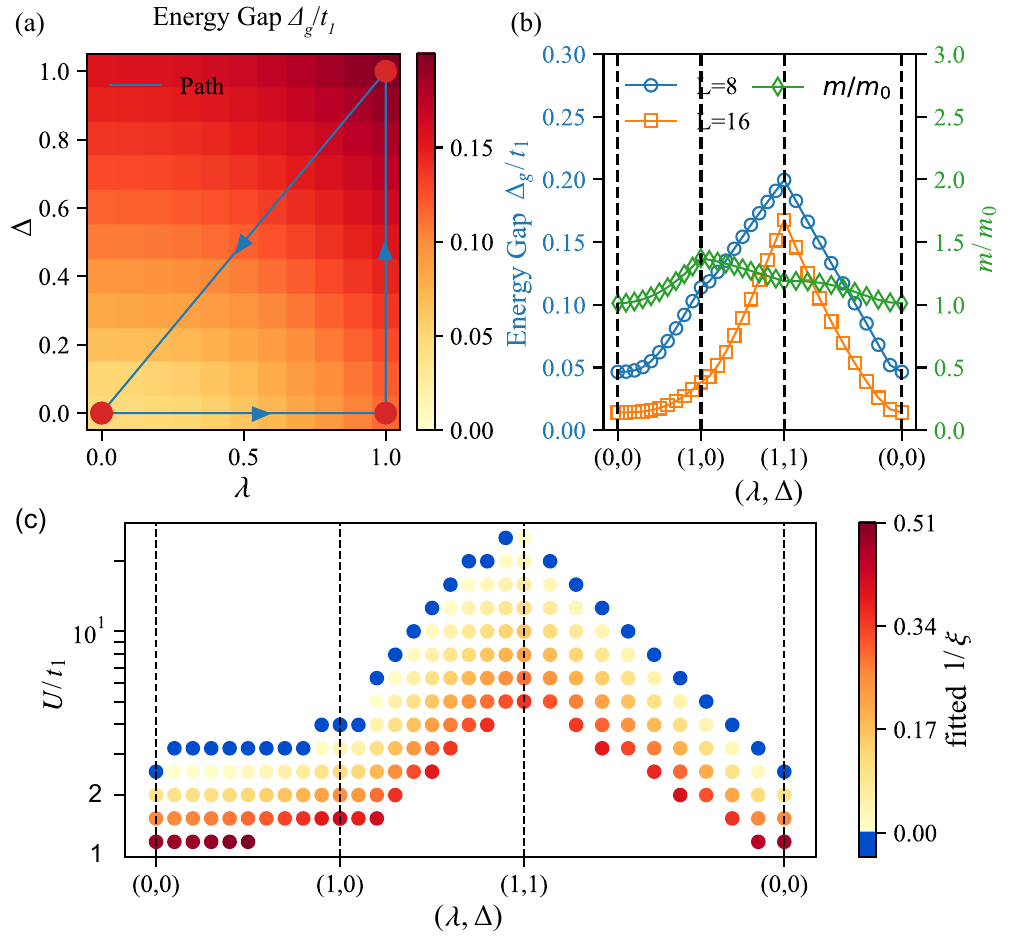}
\caption{\label{fig:4} (a) Two-hole binding gap as a function of $\lambda$ and $\Delta$ defined in Eqs.~\eqref{eq:hamilcorr} for \(U= 10t_1\). The gap shown in the heatmap is calculated from constrained ED on an $8 \times 8$ PBC cluster at momentum sector $k = (0, 2 \pi / L)$. A small shift of momentum sector is used to mitigate finite-size effects. The path for gap analysis is marked by line segments with arrows indicating the parameter sweep direction. (b) Comparison between binding gaps calculated for $8 \times 8$ and $16 \times 16$ systems along the path. The gap is smaller for the $16 \times 16$ system, and reasonable convergence is reached when the gap is large. (c) The pair correlation length $\xi$ extracted from fitting the hole-hole correlation with $|\psi|^2 = (1/r) e^{- r/\xi}$. The $x$-axis corresponds to the path, while $U/t_1$ in the vertical axis controls the strength of the attraction $J_1/2= 2 t_1^2/2$. 
While the extended $t_J$ model is not longer valid for values of $U/t_1$ or order one, we still extend the model to this unrealistic regime in order to demonstrate that the Cooper pair in absence of spin fluctuations is adiabatically connected with the one obtained in presence of spin fluctuations. The resulting coherence length for the realistic values \(U = 10 t_1\) and $\Delta=\lambda =1$ is \(\xi = 6.3\).
}
\end{figure}

Fig.~\ref{fig:4} systematically maps the evolution of the binding gap $\Delta_g$ across the two-dimensional parameter space of kinetic fluctuations ($\lambda$) and spin fluctuations ($\Delta$). Calculations are performed over $8 \times 8$ systems, and compared with results from $16 \times 16$ sites to assess finite-size convergence. Our parameter path traces from the exactly solvable limit $(\lambda, \Delta) = (0, 0)$ through the kinetically-driven regime $(\lambda, \Delta) = (1, 0)$ to the fully fluctuating $t-J-t_{\text{c-h}}$ limit $(\lambda, \Delta) = (1, 1)$. This trajectory captures the progressive enhancement of pairing as different types of fluctuations are activated. Note that we cannot reliably access the point $(\lambda, \Delta) = (0, 1)$ with our kinetically-generated basis set, as the Hilbert space truncation scheme becomes biased due to poor connectivity. And complementary large-scale DMRG simulations confirm significantly suppressed pairing tendencies in this spin-fluctuation-dominated regime (see SM).

The results demonstrate a marked enhancement of pairing correlations across the parameter space. At the exactly solvable limit $(\lambda, \Delta) = (0, 0)$, the binding gap is extremely small and decreases with system size, consistent with the expected behavior of this weakly correlated regime. Both kinetic fluctuations and spin fluctuations act synergistically to amplify pairing. The binding gap reaches its maximum value of approximately $\Delta_g = 0.19 t_1$ at $(\lambda, \Delta) = (1, 1)$, indicating the formation of tightly bound Cooper pairs. The reasonable agreement between $8 \times 8$ and $16 \times 16$ clusters, particularly in the strong pairing regime, demonstrates the reliability of our finite-size calculations.

To complement the binding gap analysis and circumvent systematic uncertainties associated with continuum edge determination, we directly extract the pair coherence length $\xi$ from the ground state wavefunction. By fitting the asymptotic decay of the hole-hole correlation function to the form: $|\psi|^2(r) \propto r^{-1} e^{-r/\xi}$, we obtain a ground state property that characterizes the spatial extent of the Cooper pairs. As shown in Fig.~\ref{fig:4}(c), the coherence length provides independent confirmation of the pairing trends observed in the binding gap.

Most importantly, our results establish an adiabatic continuity between the non-fluctuating exactly solvable limit and the full SU(2)-symmetric model. By smoothly varying the parameters $(\lambda, \Delta)$ without closing the binding energy gap, we demonstrate that the superconducting phases in both regimes share the pairing mechanism and pairing symmetry, despite their vastly different pairing strength. We verify this connection independently through detailed analysis of the superconducting pair correlations, confirming that both regimes exhibit $d_{x^2-y^2}$-wave pairing symmetry.

\section{AFM polaron enhanced hole pairing}

The significant enhancement of pairing strength with increasing $\lambda$ and $\Delta$ originates from the cooperative formation of singlet states around charge carriers. Reminiscent of the spin-bag mechanism~\cite{schrieffer_spin-bag_1988,weng_mobile_1990}, we propose an intuitional picture in which \emph{antiferromagnetic polarons} form to reduce the kinetic energy cost associated with $t_2$ hopping. Unlike the original spin-bag scenario, which is rooted in Nagaoka's theorem, these polarons are governed by the formation of singlet bonds near each hole, consistent with the ``counter-Nagaoka theorem''~\cite{Haerter2005}. When two holes occupy nearest-neighbor sites, they cooperatively enhance the singlet character along the two bonds parallel to their separation, thereby suppressing kinetic frustration. This cooperative mechanism provides a unified explanation for the enhanced pairing interactions observed with increasing $\lambda$ and $\Delta$, and proves more robust than the conventional spin-bag mechanism in the absence of kinetic frustration, as singlet formation is energetically favored by both the superexchange interaction $J_1$ and the kinetic energy.

\begin{figure}
\includegraphics[width=\columnwidth]{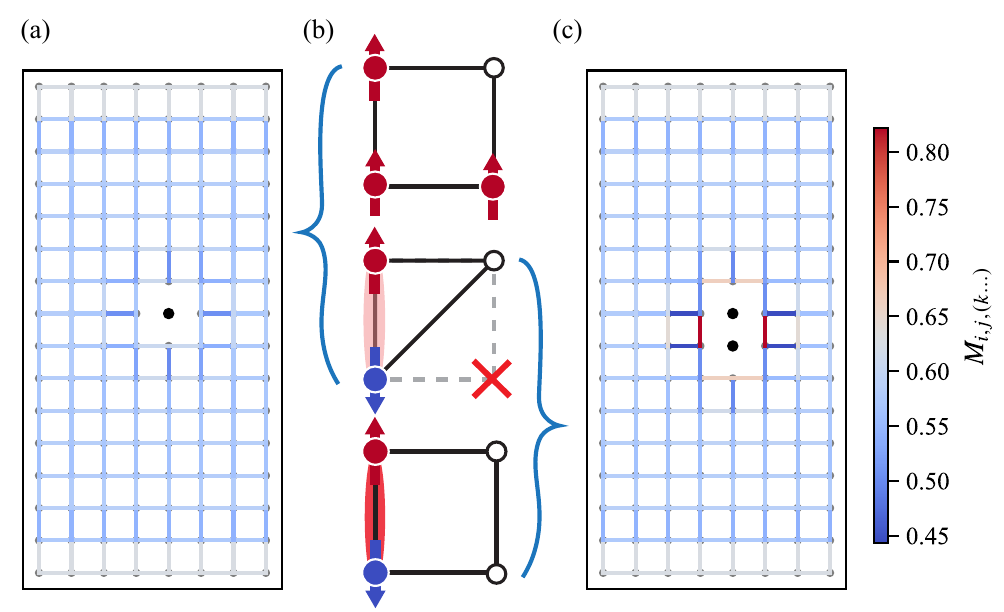}
\caption{\label{fig:5} (a) Spin-spin correlations $M_{i,j,(k)}$ between sites $i$ and $j$ with a single hole fixed at site $k$ (indicated by the black dot). The system is a single-hole-doped 8-leg cylinder of length $L_x = 16$, with parameters $t_2/t_1 = 0.6$, $\lambda = 1$, $\Delta = 1$, and $U = 10 t_1$. The DMRG calculations retain $m = 5000$ states per block. 
(b) Schematic illustration of representative kinetic processes. Black solid lines connect sites involved in each process, and shaded ellipses highlight regions of singlet formation. 
(c) Spin-spin correlations $M_{i,j,(k,l)}$ for the two-hole case, with holes fixed at sites $k$ and $l$, using the same geometry and parameter set as in panel (a).
}
\end{figure}

To quantify the AFM polaron formation, we define a conditional spin-spin correlation function:
\begin{equation}
    M_{i, j, (k, l, m \cdots)} = -\frac{\left\langle \prod_{q=k, l, m \cdots} n_{h, q}  \left({\bm S}_i \cdot {\bm S}_j - \frac14\right)\right\rangle}{\langle \prod_{q=k, l, m \cdots} n_{h, q}\rangle}
\end{equation}
which measures spin correlations between sites $i$ and $j$ when the holes are constrained to specific positions $(k, l, m, \dots)$. Fig.~\ref{fig:5}(a) shows that when a single hole is fixed at the center of the lattice, the surrounding bonds exhibit an enhanced singlet character. The fixed-hole spin-spin correlation $M_{i,j,(k)}$ increases from $0.58$—when the hole is far from the bond $ij$—to $0.61$ when $i$, $j$, and $k$ belong to the same triangle. This behavior contrasts with the suppression of singlet character observed for $t_2 = 0$, as predicted by the conventional spin-bag mechanism. Notably, as shown in Fig.~\ref{fig:5}(c), the singlet character is further enhanced to $M_{i,j,(k,l)} \simeq 0.82$ on the bonds $i,j$ parallel to the pair of neighboring holes located at $(k,l)$.

This singlet character enhancement on the parallel bonds to the Cooper pair increases the binding energy through three distinct kinetic processes. First, as shown in the top panel in Fig.~\ref{fig:5}(b), the four-site ring exchange process involving $t_1$ hoppings generates ferromagnetic correlations similar to Nagaoka mechanism. Second, three-site triangular exchange processes involving the product $t_1^2 t_2$ create KF when this product is positive (middle panel). This frustration is relieved by singlet formation, which introduces  $\pi$-flux to lower the hole KE by suppressing KF. Such frustration-induced singlet formation has been theoretically proposed and experimentally observed in triangular lattice moir\'e systems~\cite{davydova_itinerant_2023,zhang2023,foutty2023tunable,tao2024observation}.

The third and most important process occurs when two holes occupy the same square plaquette, as illustrated in the bottom panel of Fig.~\ref{fig:5}(b). While single-hole configurations tend to promote ferromagnetism via ring-exchange processes, the presence of two holes within the square cluster instead favors antiferromagnetic correlations. As a result, when two holes form a bound pair, both triangular and square hopping processes cooperatively enhance singlet formation, thereby eliminating the magnetic frustration inherent to single-hole configurations. This cooperative amplification of singlet character provides the dominant contribution to the hole-pair binding energy.

To quantify how singlet formation enhances kinetic energy, we decompose the total kinetic energy across different bonds. The kinetic energy $H_t = - \sum_{j k, \sigma} t_{jk} c_{j \sigma}^{\dagger} c_{k \sigma}$ of the two-hole system can be expressed as the expectation value of \emph{two-body} hopping processes where one hole at site $j$ moves to a neighboring site, while the other hole remains fixed at a given site $l$:
\begin{eqnarray} 
    &&\langle \Psi_G| H_t |\Psi_G\rangle= -\sum_{j, k \neq l} \tilde{t}_{j k} (l) \psi(l, j) \psi(l, k),
\end{eqnarray}
where 
\begin{equation} 
\psi(l, j) = \sqrt{\langle \Psi_G | \bar{n}_{l} \bar{n}_{ j} |\Psi_G\rangle} 
\end{equation} 
and 
\begin{equation} 
\tilde{t}_{jk} (l) = t_{j k} \frac{\langle \Psi_G| \sum_\sigma c_{j\sigma}^\dagger c_{k\sigma} \bar{n}_{l}|\Psi_G\rangle }{\psi(l, j) \psi(l, k) } 
\end{equation} 
represents the effective hopping amplitude for a transition from the hole configuration $(j, l)$ to $(k, l)$.
In this reformulation of the kinetic energy term,  $|\psi(l, j)|^2$ directly reproduces the original model's two-hole probability density, while the Cauchy–Schwarz inequality guarantees that 
\begin{equation}
|\tilde{t}_{jk} (l)| \leq |t_{jk}|.
\end{equation}
This is basically the motivation for considering the effective amplitudes $\tilde{t}_{jk} (l)$ instead of directly considering the correlation function $\langle \Psi_G| \sum_\sigma c_{j\sigma}^\dagger c_{k\sigma} \bar{n}_{l}|\Psi_G\rangle $.

\begin{figure}
    \includegraphics[width=\columnwidth]{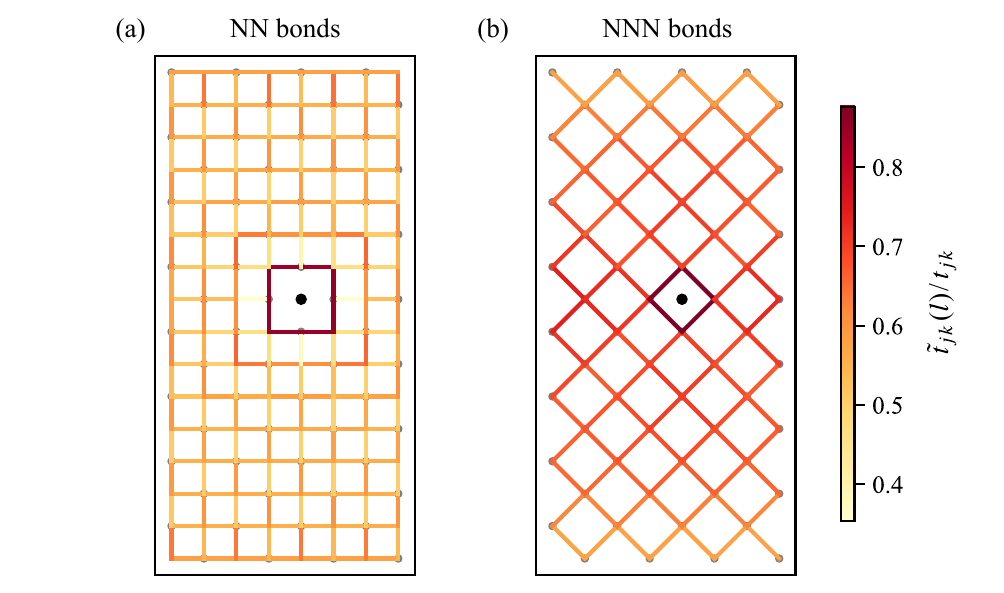}
    \caption{\label{fig:6}  (a) two body hopping amplitude \(\tilde{t}_{j k} (l)\) on NN bonds calculated using U(1) DMRG on an 8-leg cylinder with up to 8000 states per block and parameters \(\lambda = \Delta = 1, \, U=10t_1\). (b) Modified hopping amplitude on NNN bonds on opposite sublattice of fixed holes using identical computational parameters. }
    \end{figure}

Fig.~\ref{fig:6}(a, b) shows the effective two-body hopping amplitudes $\tilde{t}_{jk}(l)$ computed with DMRG for two holes on an 8-leg cylinder of length $L_x = 16$. As anticipated from Fig.~\ref{fig:5} and the suppression of kinetic frustration via enhanced singlet formation on the triangle formed by the hole and two neighboring spins, $\tilde{t}_{jk}(l)$ reaches its maximum when the hole at site $l$ lies adjacent to the bond connecting sites $j$ and $k$. In other words, as illustrated in Fig.~\ref{fig:5}(c), strong singlet correlations on the bonds parallel to the two holes suppress kinetic frustration and enhance $\tilde{t}_{jk}(l)$ when $j$, $k$, and $l$ form a minimal triangle.

In contrast, when the two holes are far apart, the singlet character around each hole weakens, as shown in Fig.~\ref{fig:5}(b), leading to less effective suppression of kinetic frustration. More quantitatively, the NN hopping amplitude is $\tilde{t}_{jk}(l) \simeq 0.85$ when the second hole is adjacent to the $jk$ bond, and it drops to $\tilde{t}_{jk}(l) \simeq 0.55$ at the largest separation allowed by the finite cluster.

A similar trend is observed for NNN hopping. When the second hole occupies the opposite sublattice, the effective amplitude is again maximized, $\tilde{t}_{jk}(l) \simeq 0.88$, when $j$, $k$, and $l$ form a minimal triangle, and minimized, $\tilde{t}_{jk}(l) \simeq 0.58$, at the largest separation allowed between the bond $jk$ and site $l$. We note that in the absence of spin fluctuations (i.e., in a perfectly classical AFM state), these two values would be identical.

A closer inspection of Fig.~\ref{fig:6} reveals that the effective hopping amplitudes $\tilde{t}_{jk}(l)$ between these two extremes exhibit a smooth oscillatory pattern. This behavior directly reflects the oscillations in the singlet character shown in Fig.~\ref{fig:5}(c), which arise from the redistribution of singlet correlations: an enhancement of singlet character on a given bond necessarily reduces it on nearby bonds due to the competition for spin singlet formation.





To verify that the polaronic mechanism—where spin fluctuations suppress kinetic frustration—provides the dominant contribution to the Cooper pair binding energy, we solved an effective two-particle problem using the two-body hopping amplitudes \(\tilde{t}_{jk}(l)\) extracted from our DMRG calculations on the 8-leg cylindrical cluster shown in Fig.~\ref{fig:6}. We assumed that these amplitudes remain unchanged beyond the largest distances accessible in the cluster (see Supplementary Material). In the thermodynamic limit, this yields a coherence length of \(\xi = 9.6\), which is reduced to \(\xi = 4.6\) when including the nearest-neighbor attractive density-density interaction arising from the Ising spin-spin interaction discussed in Section~\ref{N&P}. This result is in good agreement with the coherence length \(\xi \simeq 6.3\) obtained from our truncated ED calculations on the largest accessible \(16 \times 16\) system, using the same Hamiltonian parameters \((\lambda = \Delta = 1,\, U = 10 t_1)\). In other words, the attractive interaction generated by the polaronic mechanism through two-body hopping is sufficient to explain a coherence length of only a few lattice spacings.

Importantly, the alternative source of attraction associated with the Ising spin-spin interaction leads to a coherence length several orders of magnitude larger than \(\xi\). This stark contrast supports the conclusion that the polaronic effect is the dominant pairing mechanism and constitutes the primary ``glue'' binding the two holes into a Cooper pair.

\section{Avoidance of phase separation}
\label{sec:aps}

\begin{figure}
\includegraphics[width=\columnwidth]{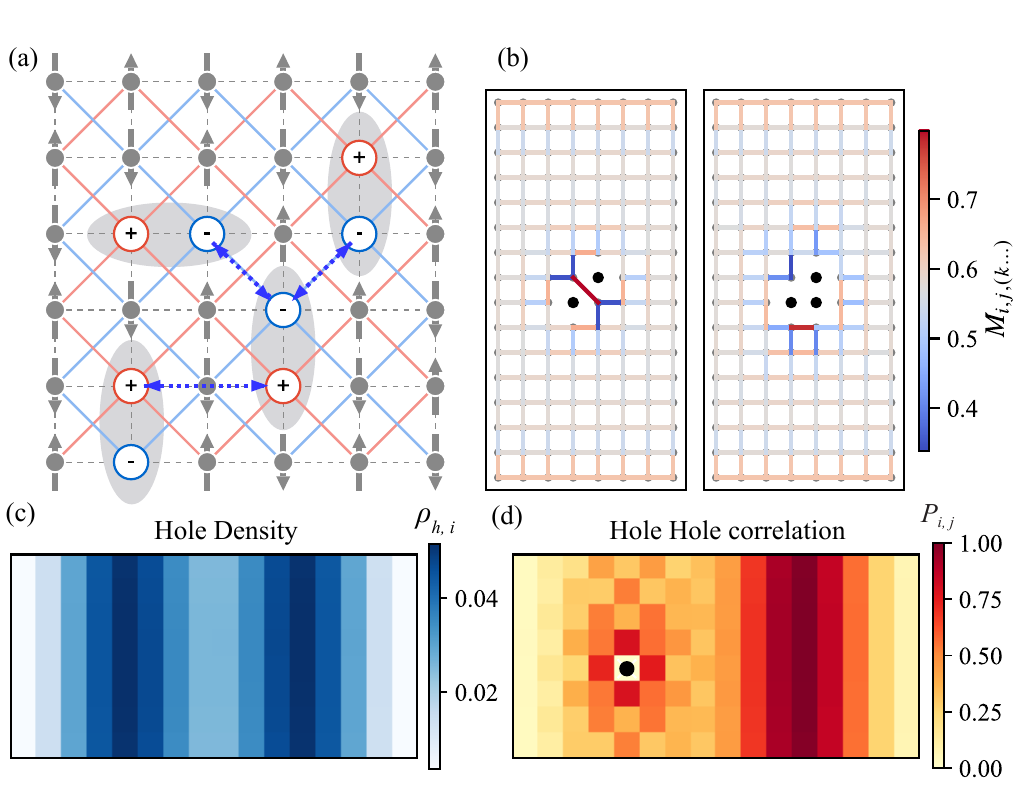}
\caption{\label{fig:7} (a) Schematic diagram illustrating the suppression of phase separation suppression from sublattice-dependent interaction, where holes (empty circles) carry sublattice labels ($+$ and $-$ symbols). The thick dashed blue line with outward arrows represents repulsion between holes on the same sublattice. (b) Spin correlation patterns when two/three holes cluster together with two holes on the same sublattices. Weakened singlet formation and enhanced triplet tendencies on adjacent bonds are shown. Breaking of symmetry in correlation pattern is caused by lattice geometry. (c) Hole density distribution $\rho_{h, i} = \langle\Psi_G | \bar{n}_{i}|\Psi_G\rangle$ for four holes calculated using DMRG with $\lambda = \Delta = 1$, $U = 10 t_{1}$, and maximum bond dimension of 5000, revealing two spatially separated Cooper pairs. (d) Corresponding hole-hole correlation function $P_{i,j} = \langle\Psi_G | \bar{n}_{i} \bar{n}_{j} |\Psi_G\rangle$ displaying both short-range pairing and long-range inter-pair repulsion.}
\end{figure}

Having established the origin of the formation of Cooper pairs, we now address the competing instabilities that typically inhibit superconductivity. Phase separation~\cite{Emery1990, Moreo1991} and charge density waves (stripes)~\cite{zheng2017stripe,huang2017numerical,jiang2019superconductivity} are among the primary competitors to uniform superconducting states. The polaronic mechanism described above, which suppresses kinetic frustration, naturally avoids phase separation. Holes experience an effective attraction when they reside on opposite sublattices [see Fig.~\ref{fig:7}(a)], and the ``glue'' binding the two holes arises from singlets formed on the bonds parallel to the hole pair, as illustrated in Fig.~\ref{fig:5}(c). Importantly, the interaction between a preformed pair and a third hole is effectively repulsive, since the additional hole would necessarily disrupt one of the singlets that stabilizes the pair.

Fig.~\ref{fig:7}(b) highlights the reduction in singlet character when three holes cluster together. Compared to an isolated pair of adjacent holes, singlet correlations on nearby bonds become weaker and more spatially delocalized, with some bonds even developing enhanced triplet character. This redistribution increases the kinetic energy, signaling an effective repulsion between the original pair and the additional hole, thereby making such clustered configurations energetically unfavorable.

Moreover, three-hole clusters inevitably place two holes on the same sublattice, suppressing the $t_2$-hopping process on at least one bond and further reinforcing the effective three-body repulsion. Since the magnetic polaron distortion remains localized around the holes, this repulsive interaction is short-ranged and prevents the formation of macroscopic phase separation or Wigner crystallization.

To test this sublattice-dependent interaction mechanism under realistic conditions, we performed DMRG simulations with four holes on a cylindrical geometry. Fig.~\ref{fig:7}(c,d) confirm our theoretical expectations: the ground state exhibits two well-separated Cooper pair wavepackets localized near opposite ends of the cylinder. The hole-hole correlation function displays the same short-range pairing signature as in the two-hole case, along with a long-range peak indicative of inter-pair repulsion. This real-space segregation of hole pairs provides direct numerical evidence for repulsive interactions between Cooper pairs, thereby supporting our proposed mechanism for avoiding phase separation.

Having established both strong pairing and phase separation suppression, we estimate superconducting transition temperatures using our calculated binding gaps.
For sufficiently low hole concentration $\rho_h$, such that $\xi \rho_h^{1/2} \ll 1$, the system enters the Bose-Einstein condensation (BEC) regime, where superconductivity arises from the superfluid transition of preformed pairs. In this limit, the critical temperature depends only on the superfluid density $\rho \simeq \rho_h/2$ and the effective mass of the pairs $m_b$, and is given by $k_B T_c \approx C \frac{\hbar^2 \rho}{m_b}$, where { $C \approx \frac{2\pi}{\log(380)}$ is the contact repulsive pair-pair interaction in the continuum (long wavelength) limit. Hole concentration level and weak interactions modify the constant $C$ only slightly though a double logarithmic dependence \cite{Nikolay2001}.} For the full $t$-$J$ model with $(\lambda, \Delta) = (1, 1)$, our spectrum for a $16 \times 16$ cluster with PBC yields $k_B T_c \approx 1.16 \, \rho_h \, t_1$ in the BEC regime.

\section{Conclusions}

In this work, we identify kinetic frustration as a straightforward and effective mechanism for generating tightly bound Cooper pairs in doped antiferromagnets. This approach diverges from earlier efforts that focused on exchange frustration to stabilize superconductivity~\cite{jiang_high_2021,jiang2023superconducting}. Specifically, we explore how the inclusion of $t_2$ stabilizes AFM correlations and promotes singlet formation around the doped holes, since $t_2$ hopping amplitude competes with $t_1$ in a uniform FM background. As elaborated in the preceding sections, this seemingly simple phenomenon gives rise to coherent $d$-wave superconductivity featuring tightly bound Cooper pairs with a correlation length of approximately 6 lattice spacings ($\xi \simeq 6$). Our finding aligns with DMRG studies of strong $d$-wave superconductivity on electron-doped square lattice Hubbard model~\cite{jiang_high_2021,chen2025global}.

To isolate the dominant contribution to the strong pairing mechanism, we began by considering a spin-fluctuation-free model in which the \(J_z\) exchange induces an effective attraction between holes with \(d_{x^2 - y^2}\) symmetry. In this limit, holes naturally reside on opposite sublattices. Upon introducing spin fluctuations via the \(t_1\) hopping and transverse exchange interactions \(J_{x,y}\), we observe the emergence of local singlet formation around the charge carriers. The singlets are formed around each hole to generate effective $\pi$-fluxes on the elementary trimers containing the hole and the corresponding two spins. To account for this effect, we employed a constrained ED approach, which we benchmarked against DMRG calculations.

The constrained ED approach systematically incorporates spin fluctuations around holes to capture polaronic effects~\cite{trugman_interaction_1988, trugman_spectral_1990}. Earlier studies applied this method to the standard \(t\)\nobreakdash-\(J\) model with nearest-neighbor hopping (\(t_1\) only), considering only configurations generated by hole motion and assuming a classical antiferromagnetic background~\cite{trugman_interaction_1988, trugman_spectral_1990}. This assumption is even more appropriate in the \(t_1\)\nobreakdash-\(t_2\)\nobreakdash-\(J_1\)\nobreakdash-\(J_2\)-$H_{\text{c-h}}$ model studied here, where the AFM phase is stabilized not only by the exchange interaction \(J_1\), but also by kinetic frustration introduced by the next-nearest-neighbor hopping \(t_2\). More importantly, because the variational ED method captures the spin fluctuations that dress each hole (and the hole pair), it provides an ideal framework for highlighting the key features of the dominant pairing mechanism.

An important advantage of the constrained ED approach is that it enables simulations on clusters as large as \(16 \times 16\) sites, significantly exceeding the maximum sizes accessible via DMRG. The DMRG results were primarily used to cross-validate the main conclusions drawn from the constrained ED calculations. In particular, both methods consistently identify spin-fluctuation-induced singlet formation as the primary mechanism driving unconventional \(d_{x^2 - y^2}\) superconductivity.

In essence, the ``bipolaron'' formed around a hole pair enables effective two-body hopping through multi-\(\pi\)-flux generation, which is strong enough to bind the two holes. This mechanism is, in some sense, analogous to the spin-bag scenario proposed by Schrieffer for the unfrustrated case~\cite{schrieffer_spin-bag_1988, Schrieffer89}, where kinetic energy promotes ferromagnetic correlations in accordance with Nagaoka's theorem~\cite{nagaoka_ferromagnetism_1966}. A key difference, however, lies in the energy scales: the spin-bag mechanism requires very large values of \(U/t_1\) for polarons to form with appreciable size, since ferromagnetic polarization is strongly penalized by the antiferromagnetic exchange \(J_1\). In contrast, the antiferromagnetic polaron emerging in our model has a relatively large spatial extent even for realistic values of \(U/t_1 \simeq 10\), simply because both the kinetic energy and \(J_1\) cooperatively favor singlet formation. The polaron arises naturally because \(J_1\) favors a \emph{uniform} distribution of singlet character over all bonds, while the kinetic energy drives an enhancement of singlet correlations in the vicinity of each hole. While kinetic frustration accounts for the majority of pairing enhancement, the AFM exchange coupling $J_1$ provides the remaining contribution.



As explained in Sec.~\ref{sec:aps}, an appealing aspect of this novel pairing mechanism is that it naturally avoids phase separation due to an effective repulsive three-body interaction. Specifically, the presence of a third hole near a Cooper pair increases the total kinetic energy, thereby disfavoring clustering. Another distinguishing feature of this mechanism is the formation of two neighboring bonds with enhanced singlet character surrounding each Cooper pair. As discussed in the introduction, the accumulation of such singlet-rich regions is expected to drive a quantum phase transition from antiferromagnetic order to a quantum paramagnetic state when the bipolarons begin to overlap significantly (noting that the linear extent of a bipolaron -comprising both the Cooper pair and the surrounding magnetic distortion- generally exceeds the coherence length \(\xi\)). The resulting state is characterized by a spin gap and short-range magnetic correlations. 

Our theoretical framework aligns with experimental observations across multiple families of unconventional superconductors, where \(d\)-wave superconductivity coexists with enhanced short-range singlet correlations~\cite{Kastner1998,Lee2006,Keimer2015,Stock2008,Leiner2014}. 
While similar phenomenology previously motivated proposals by P.~W.~Anderson and others for doping the parent resonating valence bond (RVB) states as the origin of high-\(T_c\) superconductivity~\cite{baskaran_resonating_1987,anderson_resonating_1987,anderson_physics_2004}, our work presents a microscopic picture in which doped charges facilitate spin singlets formation and a finite critical doping is required to destabilize the antiferromagnetic parent state. Whether the resulting quantum paramagnet can be continuously connected to an RVB-like wave function remains an open and compelling question for future investigation.

\begin{acknowledgments}

We are grateful to Collin Broholm, Philip Philips, Donna Sheng and Hongcheng Jiang for helpful discussions. C.D.B. acknowledges support from the U.S. Department of Energy, Office of Science, Office of Basic Energy Sciences, under Award Number DE-SC0022311. 
\end{acknowledgments}

\bibliography{apssamp}

\clearpage 
\appendix
\onecolumngrid

\section*{Supplementary material for: Antiferromagnetism and Tightly Bound Cooper Pairs Induced by Kinetic Frustration}

\setcounter{section}{0}
\setcounter{figure}{0}
\setcounter{equation}{0}
\renewcommand{\thefigure}{S\arabic{figure}}

\twocolumngrid

\section{The Kinetically stabilized Néel order}

In this section, we discuss the emergence of antiferromagnetic Néel order in the square-lattice Hubbard model through kinetic energy minimization in the infinite U limit. Analogous to Nagaoka ferromagnetism—where itinerant electrons stabilize ferromagnetic order via kinetic constraints—antiferromagnetic Néel order arises here without explicit magnetic exchange interactions. The mechanism hinges on kinetic frustration: destructive quantum interference between degenerate electron hopping pathways energetically penalizes uniform spin configurations, favoring a staggered arrangement. The resulting kinetically driven Néel order provides the microscopic foundation for AFM polaron formation and the pairing mechanism discussed in following sections.

\subsection{Sublattice ferromagnetism through Nagaoka mechanism at $t_1=0$}

In the $t_1 = 0$ limit, the $t_2$ terms mediate hopping exclusively within each sublattice (indicated in red and blue in Fig.~\ref{fig:2}), effectively decoupling the system into two independent subsystems. Upon doping two holes into the half-filled system, the ground state $|\Psi_G\rangle$ takes the form:
\begin{equation}
    |\Psi_G\rangle = \sum_{\bm{r}_1, \bm{r}_2} \psi(\bm{r}_1, \bm{r}_2)|\phi(\bm{r}_1, \bm{r}_2)\rangle \label{decompose wavefunction with holes}
\end{equation}
where $\psi(\bm{r}_1, \bm{r}_2)$ are superposition coefficients for configurations with holes at positions $\bm{r}_1$ and $\bm{r}_2$, and $|\phi(\bm{r}_1, \bm{r}_2)\rangle$ denotes a many-body state with holons at sites $\bm{r}_1, \bm{r}_2$ and arbitrary spin configurations on the remaining sites. These states satisfy the orthonormality condition $\langle\phi(\bm{r}_1, \bm{r}_2)|\phi(\bm{r}_1', \bm{r}_2')\rangle = \delta_{\bm{r}_1, \bm{r}_1'} \delta_{\bm{r}_2, \bm{r}_2'}$, and the wavefunction is normalized such that $\langle\Psi_G|\Psi_G\rangle=1$.

The ground state energy is bounded by:
\begin{eqnarray}
    E_0 = \langle\Psi_G|\mathcal{H}|\Psi_G\rangle &\geq& - 2 t_2  \sum_{\bm{r}_1, \langle\langle \bm{r}_2, \bm{r}_3\rangle\rangle} |\psi(\bm{r}_1, \bm{r}_2)^* \psi(\bm{r}_1, \bm{r}_3)| \nonumber\\
    &\geq& - 4 t_2 \sum_{\bm{r}_1, \bm{r}_2} |\psi(\bm{r}_1, \bm{r}_2)|^2 = - 8 t_2
\end{eqnarray}

The energy is minimized under the condition:
\begin{equation}
    \psi(\bm{r}_1, \bm{r}_2) = \frac{2}{N} (-1)^{(x_1 + x_2 - 1)}
\end{equation}
for $\bm{r}_1, \bm{r}_2$ located on different sublattices, and zero otherwise. Here, \( x_1 \) and \( x_2 \) denote the \( x \)-coordinates of \( \bm{r}_1 \) and \( \bm{r}_2 \), respectively. The optimal spin background features uniform spin alignment within each sublattice, while the relative orientation between the two sublattices remains arbitrary:
\begin{equation}
    |\phi(\bm{r}_1, \bm{r}_2)\rangle = c_{\hat{\bm{n}}_1, \bm{r}_1} c_{\hat{\bm{n}}_2, \bm{r}_2} |\hat{\bm{n}}_1 \hat{\bm{n}}_2\rangle
\end{equation}
where the background spin state is:
\begin{equation}
    |\hat{\bm{n}}_1 \hat{\bm{n}}_2\rangle = \prod_{i \in A} c^{\dagger}_{\hat{\bm{n}}_1, i} \prod_{j \in B} c^{\dagger}_{\hat{\bm{n}}_2, j} |0\rangle
\end{equation}

Here, $c^{\dagger}_{\hat{\bm{n}}, i}$ creates a fermion at site $i$ with spin pointing along unit vector $\hat{\bm{n}}$. The state $|\hat{\bm{n}}_1 \hat{\bm{n}}_2\rangle$ corresponds to a spin configuration in which all spins on sublattice A align along $\hat{\bm{n}}_1$, while those on sublattice B align along $\hat{\bm{n}}_2$.

\subsection{Antiferromagnetic order through inter-sublattice hopping $t_1$}

We now discuss how a small nearest-neighbor hopping amplitude $t_1$ perturbatively lifts the degeneracy among different relative spin orientations of the two sublattices. This term introduces a coupling that energetically favors specific spin alignments, thereby stabilizing antiferromagnetic order.

In the ferromagnetic configuration ($\hat{\bm{n}}_1 = \hat{\bm{n}}_2 = \hat{\bm{z}}$), where all spins are uniformly polarized, the $t_1$ hopping processes between sublattices interfere destructively, causing the probability amplitude for a hole to hop to the opposite sublattice to cancel exactly. As illustrated in Fig.~\ref{fig:neel_order_parameter.pdf}(a), when a hole attempts to move across sublattices, the resulting amplitudes acquire opposite signs due to the phase factor $(-1)^{(x_1 + x_2 - 1)}$ in the wavefunction. This destructive interference ensures that the original $t_2$-only ground state remains an exact eigenstate, even in the presence of a finite $t_1$, as long as the system remains fully spin-polarized.

However, when the two sublattices are polarized in opposite spin directions, the spin mismatch left behind by a hole’s motion differs between the various $t_1$ hopping paths. This asymmetry removes the exact cancellation described previously and enables the system to lower its energy via second-order processes—namely, a hole hops to the opposite sublattice  and then returns. These second-order virtual hoppings respect the underlying $SU(2)$ symmetry and couple only sublattice spin states with magnetic quantum numbers $m$ and $m \pm 1$. As a result, the corresponding energy reduction takes the following form:
\begin{equation}
    \Delta E = c (\bm{S}_1 \cdot \bm{S}_2 - \bm{S}_1^2)
\end{equation}
Here, $\bm{S}_1$ and $\bm{S}_2$ denote the total spins of the two sublattices, with equal magnitude, and $c > 0$ is a positive constant. This implies that the lowest-energy state resides in the $S_{\text{total}} = 0$ sector, where $\bm{S}_1$ and $\bm{S}_2$ are aligned antiparallel to form a singlet ground state.

The Néel order parameter or staggered magnetization that emerges in the thermodynamic limit can be extrapolated from a finite size scaling study of the square of the staggered magnetization normalized by the total number of sties $N_{\mathrm{s}}$:
\begin{equation}
    m^2(\boldsymbol{q})=\frac{1}{N_{\mathrm{s}}^2} \sum_{i, j} e^{i \boldsymbol{q} \cdot\left(\boldsymbol{r}_i-\boldsymbol{r}_j\right)}\left\langle\Psi_G|\boldsymbol{S}_i \cdot \boldsymbol{S}_j|\Psi_G\right\rangle
\end{equation}
with \(\boldsymbol{q} = (\pi, \pi)\).  For the ground state singlet wavefunction in $S_{\text{total}} = 0$ sector, we have  $ m^2(\boldsymbol{q})= (1/4 - 1/N^2_{\mathrm{s}} ) $, which approaches 0.25  in the thermodynamic limit.

\subsection{Robustness of the kinetically stabilized Néel order for finite $t_1/t_2$}

\begin{figure}
\includegraphics[width=\columnwidth]{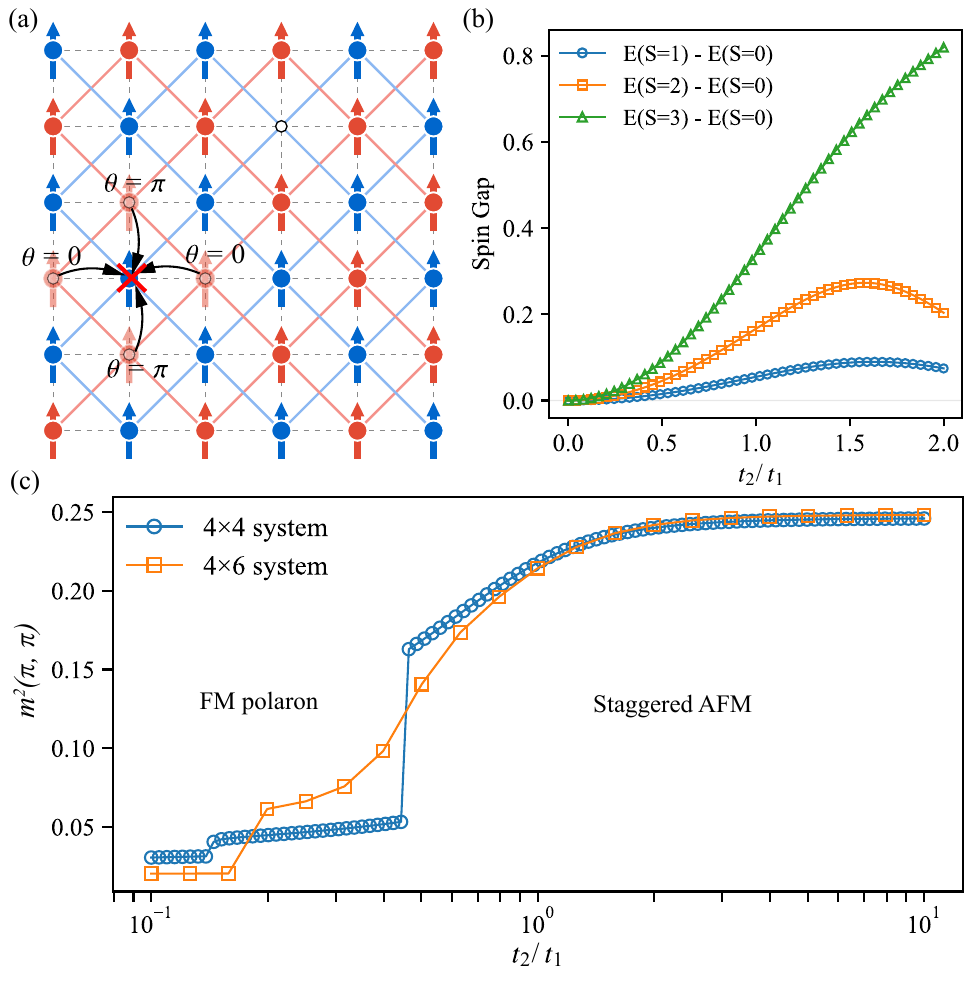}
\caption{\label{fig:neel_order_parameter.pdf} (a) Schematic diagram for the cancellation of hopping processes between opposite sublattices. The phase of superposition coefficient for $|\Psi_{m_1 = m_2 = J}\rangle$ is marked on the graph. (b) Spin gaps for each total spin sector calculated in a $4 \times 4$ PBC cluster. (c) Square of the  Néel order parameter: $m(\boldsymbol{q})^2=\frac{1}{N_{\mathrm{s}}^2} \sum_{i, j} e^{i \boldsymbol{q} \cdot\left(\boldsymbol{r}_i-\boldsymbol{r}_j\right)}\left\langle\boldsymbol{S}_i \cdot \boldsymbol{S}_j\right\rangle$ with respect to \(t_2/t_1\) calculated from ED of $4 \times 4$ and $4 \times 6$  clusters with PBC. }
\end{figure}

On top of the perturbative wavefunction approach, we now verify the stability of the kinetically stabilized Néel order at large $t_1$ regime through numerical exact diagonalization using the XDiag and QuSpin packages~\cite{Wietek2025, Weinberg2017a, weinberg_quspin_2019}. Fig.~\ref{fig:neel_order_parameter.pdf}(b) shows the ground state remains in the $S_{\text{total}} = 0$ sector even as $t_2/t_1$ down to values around 0.5. The Néel order parameter there maintains values close to the maximum 0.25 from $t_2/t_1 \to \infty$ down to $t_2/t_1 \sim 0.5$ [Fig.~\ref{fig:neel_order_parameter.pdf}(c)]. This result is consistent with large Néel order parameter when a single hole is doped under intermediate $t_2/t_1$ \cite{sposetti_classical_2014}. 

For our chosen parameter ratio $t_2/t_1 = 0.6$, the Néel order parameter reaches approximately 0.17, indicating strong antiferromagnetic correlations with relatively suppressed spin fluctuations. This robust magnetic order provides a stable parent state for doping induced antiferromagnetic polarons.

\section{ Exactly solvable limit \(\lambda=\Delta=0\)   \label{sec: lambda delta 00 limit}}

\subsection{Hamiltonian and wavefunction structure}

Building on our previous kinetic magnetic order analysis in the infinite-$U$ limit with small $\lambda$, we now turn to the exactly solvable limit $\lambda = \Delta = 0$. In this regime, the nearest-neighbor hopping $t_1$ vanishes, and the Hamiltonian features Ising-type spin interactions:
\begin{eqnarray}
H &=& -t_2 \sum_{\langle\langle j l\rangle\rangle, \sigma} c_{j \sigma}^{\dagger} c_{l \sigma} + h.c.
+ \frac{J_{1}}{2} \sum_{\langle j l\rangle}  \left[S^z_j  S^z_l -\frac{n_j n_l}{4} \right] \nonumber \\
&&+ \frac{J_{2}}{2} \sum_{\langle\langle j l\rangle\rangle}  \left[S^z_j  S^z_l -\frac{n_j n_l}{4} \right],
\label{t2-Jz}
\end{eqnarray}
where $\langle j l\rangle$ and $\langle\langle j l\rangle\rangle$ denote nearest-neighbor and next-nearest-neighbor pairs, respectively.

The hole-hole interaction arises solely from differences in Ising interaction energies between adjacent and isolated hole configurations. As detailed in the main text, when two holes occupy neighboring lattice sites, the number of broken antiferromagnetic bonds is reduced from eight to seven. This reduction effectively generates an attractive interaction of magnitude $J_1/2$ between  adjacent holes.

In the absence of $t_1$, holes are confined to their respective sublattices, each thus acquiring a fixed sublattice index. Consequently, the two-hole ground state wavefunction can be expressed as:
\begin{equation}
    |\Psi_G\rangle = \sum_{\bm{r}_1 \in A, \bm{r}_2 \in B} \psi(\bm{r}_1, \bm{r}_2) c_{1 \uparrow}  c_{2 \downarrow}|\uparrow\downarrow...\uparrow\downarrow\rangle
\end{equation}

Here, $c_{1\uparrow}$ and $c_{2 \downarrow}$ are annihilation operators that annihilate electrons at positions $\bm{r}_1$ (on sublattice A) and $\bm{r}_2$ (on sublattice B) with spins up and down, respectively. The background state $|\uparrow\downarrow...\uparrow\downarrow\rangle$ represents the classical Néel antiferromagnet with up spins on sublattice A and down spins on sublattice B, which is the ground state for the half-filled Hamiltonian in Eq. \ref{t2-Jz}.

For translationally invariant systems, we can fix the center-of-mass momentum to $\bm{K}$ and rewrite the wavefunction in terms of the center-of-mass coordinate $\bm{R} = \frac{\bm{r}_1 + \bm{r}_2}{2}$ and relative coordinate $\bm{\rho} = \bm{r}_1 - \bm{r}_2$:
\begin{equation}
    \psi(\bm{r}_1, \bm{r}_2) = \tilde{\psi}_{\bm{\rho}} e^{i \bm{K} \cdot  \bm{R}}
\end{equation}

The relative wavefunction $\tilde{\psi}_{\bm{\rho}}$ satisfies the equation:
\begin{equation}
    \sum_j t_{jl}\cos\left( \frac12 \bm{K} \cdot (\bm{\rho}_j- \bm{\rho}_l)\right)\tilde{\psi}_{\bm{\rho}_j} - \frac{J_{1}}{2} \delta(|\bm{\rho}_j|- a)
    \tilde{\psi}_{\bm{\rho}_l} =  E \tilde{\psi}_{\bm{\rho}_l}
    \label{eq:fix momentum equation}
\end{equation}
where $t_{ij} = t_2$ for next-nearest neighbors and 0 otherwise, and $a$ is the lattice constant.

\subsection{Pairing symmetry and Cooper pair wavefunction}

Analyzing the two-hole Cooper pair wavefunction yields essential insights into the pairing symmetry and spatial structure in the $(\lambda, \Delta) = (0, 0)$ limit. By solving the eigenvalue equation Eq.~\eqref{eq:fix momentum equation} for various center-of-mass momenta $\bm{K}$, we can clearly characterize the underlying pairing mechanism in this exact solvable limit.

Our analysis confirms that the global energy minimum occurs in $\bm{K}=(0, 0)$ sector, where the Hamiltonian is real and time-reversal invariant. The corresponding bound state has $d_{x^2-y^2}$-wave pairing symmetry emerging from the positive tunneling $t_2$ which leads to a sign alternating sign alternation of one hole moving around the other one. In other words, the $d_{x^2-y^2}$ symmetry of the pair is a direct consequence of the kinetic energy frustration.  

Fig.~\ref{fig:three_panel_wavefunction_scaling.pdf} illustrates how the Cooper pair wavefunction $\tilde{\psi}_{\bm{\rho}}$ changes with the interaction strength $U$. Here we include values of $U$ for which the original $t-J$ model is obviously invalid, as a way to tune the effective NN attractive interaction estimating the magnitude of the attraction that would be required to obtain a coherence length $\xi$ of order a few lattice spaces.  Panel (a) shows the wavefunction for $U = 10t_1$ (weak exchange with $J_1 \approx 0.4t_1$), while panel (b) shows it for $U = 2t_1$ (strong exchange with $J_1 \approx 2t_1$). The $d_{x^2-y^2}$ symmetry is evident from the sign changes (color transitions between red and blue) along different directions. 

As expected, $\xi$ varies dramatically with interaction strength. For large $U$ (panel a), the slow spatial decay of the wavefunction indicates a weakly bound state characterized by  a coherence length $\xi$ several orders of magnitude higher than $a$. Conversely, at smaller $U$ (panel b), the wavefunction decays rapidly with distance, revealing a tightly bound pair with a coherence length $\xi$ of only a few lattice spacings.

\subsection{Continuum limit analysis and binding energy}

To further understand the pairing structure, we consider the continuum limit analysis, which reveals crucial properties of bound states in the $(\lambda, \Delta) = (0, 0)$ limit. The continuum approximation is particularly effective when the Cooper pair wavefunction varies slowly over multiple lattice sites—precisely the scenario encountered at small $J_{1}/t_2$ ratios, where the coherence length significantly exceeds the lattice spacing ($\xi \gg a$).

In this limit, for widely separated holes, we transition our discrete lattice model to a continuum description, expressing the relative wavefunction as:
\begin{equation}
    \tilde{\psi}_{\bm{\rho}_i} = (-1)^{(X_i - 1)} \sqrt{2} a \tilde{\phi}_{\bm{\rho}_i}
\end{equation}
Here, $\tilde{\phi}_{\bm{\rho}}$ represents a smoothly varying function, while the oscillating factor $(-1)^{(X_i - 1)}$ with \(X_i\) denoting the \( x \)-coordinate of \( \bm{\rho}_i \) captures the inherent $d$-wave sign alternation. The prefactor $\sqrt{2}a$ normalizes the wavefunction by effective lattice constant of each sublattice to ensure \(\int |\tilde{\phi}_{\bm{\rho}}|^2 d\rho^2 = 1\) in the continuum limit.

The resulting effective Hamiltonian in the continuum limit comprises two main components: a kinetic term from $t_2$ hopping: $-8t_2 - 4t_2a^2\int \tilde{\phi}(\bm{\rho})^*\nabla^2\tilde{\phi}_{\bm{\rho}}\,d\rho^2$ and a potential term originating from the nearest-neighbor Ising interaction: $v_0|\tilde{\phi}(0)|^2$ where $v_0 = 4a^2J_{1}$

The constant $-8t_2$ represents the ground state energy of two non-interacting holes, while the gradient term describes relative motion with effective mass $m^* = 1/8(\hbar/a)^2t_2$—exactly half the single-particle mass, as expected for the reduced mass in a two-body problem.

The continuum Schrödinger equation thus becomes:
\begin{equation}
    -\frac{\hbar^2}{2m^*}\nabla^2\tilde{\phi}_{\bm{\rho}} - v_0\delta(\bm{\rho})\tilde{\phi}_{\bm{\rho}} = (E + 8t_2)\tilde{\phi}_{\bm{\rho}}
\end{equation}

While the true interaction has finite width rather than being a delta function, this distinction affects only the coefficient of the binding energy without changing its scaling behavior.

A key result from quantum mechanics establishes that in two dimensions, even infinitesimally weak attractive potentials yield bound states with binding energies that scale as~\cite{deLlano2003}:
\begin{equation}
    \Delta_g = E_1 - E_0 \sim \exp\left(-\frac{2\hbar^2\pi}{m^*v_0}\right) \sim \exp\left(-\frac{4\pi|t_2|}{J_{1}}\right)
\end{equation}

This exponential relationship explains the dramatic binding strength variation with $U$. At large $U$ (small $J_{1}$), binding remains extremely weak, but strengthens considerably as $U$ decreases (increasing $J_{1}$).

The bound-state wavefunction correspondingly takes the form at \(|\rho|\gg 1\):
\begin{equation}
    \tilde{\phi}_{\bm{\rho}} \sim K_0\left(\frac{\sqrt{2m^*\Delta_g}}{\hbar} \rho\right)
\end{equation}
where $K_0$ is the modified Bessel function of the second kind of order zero and $\rho \equiv |\bm{\rho}|$.

For distances where $\frac{\sqrt{2m^*\Delta_g}}{\hbar}\rho \gg 1$, the wavefunction assumes the asymptotic form:
\begin{equation}
    \tilde{\phi}_{\bm{\rho}} \sim \frac{1}{\sqrt{\rho}}e^{-\frac{\sqrt{2m^*\Delta_g}}{\hbar}\rho}
    \label{eq:exponential decay}
\end{equation}

\begin{figure}
\includegraphics[width=\columnwidth]{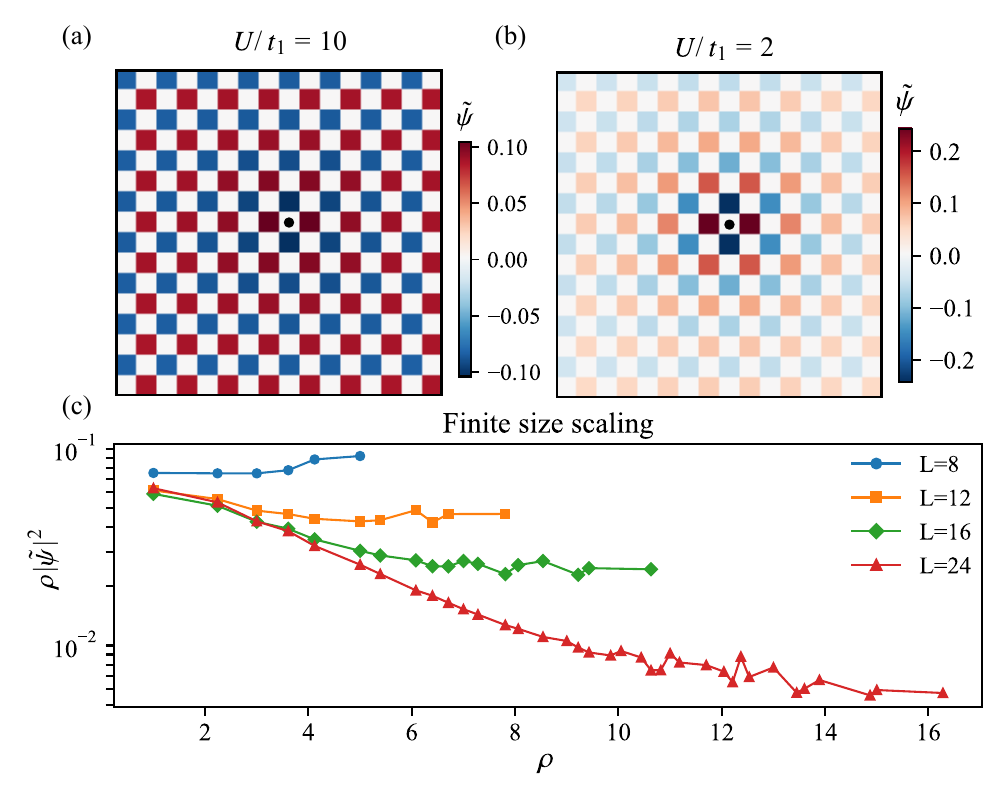}
\caption{\label{fig:three_panel_wavefunction_scaling.pdf} 
(a) Wavefunction $\tilde{\psi}_{\bm{\rho}}$ of the relative coordinate of the Cooper pair for $U = 10t_1$ and $\lambda = \Delta = 0$. The black dot marks the origin. A clear $d_{x^2 - y^2}$-wave symmetry is observed. 
(b) Same as in (a), but for $U = 2t_1$. 
(c) Scaling of $\log\left[\rho |\tilde{\psi}_{\bm{\rho}}|^2\right]$ as a function of $\rho$. A linear dependence is seen at large distances, away from the system boundaries. Smaller system sizes tend to yield incorrect scaling behavior.
}
\end{figure}

This expression directly connects the binding energy $\Delta_g$ to the Cooper pair coherence length 
\begin{equation}
\xi = \frac{\hbar}{\sqrt{2m^*\Delta_g}} \sim e^{\frac{2\pi|t_2|}{J_{1}}}.
\end{equation}
$\xi$ grows exponentially in $|t_2|/J_1$, explaining the nearly uniform amplitude observed in Fig.~\ref{fig:three_panel_wavefunction_scaling.pdf}(a).

\subsection{Finite size effects}

Finite-size effects significantly impact both the spatial profile of Cooper pairs and their binding energies in numerical simulations. Understanding these effects is crucial for correctly interpreting our numerical results and extrapolating to the thermodynamic limit.

Two key observables exhibit pronounced finite-size sensitivity. 

First, the spatial structure of the Cooper pair wavefunction is strongly affected. Fig.~\ref{fig:three_panel_wavefunction_scaling.pdf}(c) shows that periodic boundary conditions (PBC) artificially enhance hole-hole correlations, particularly in small systems where holes can interact with their periodic images. This artifact can be severe enough to reverse the sign of the slope when fitting correlation functions to Eq.~\eqref{eq:exponential decay}, leading to the misleading appearance of repulsion rather than attraction.

Second, the calculation of the binding energy requires careful treatment of finite-size effects. The conventional definition, $\Delta_g[\bm{k}=(0,0)] = E_1[\bm{k}=(0,0)] - E_0[\bm{k}=(0,0)]$, includes spurious finite-size contributions arising from the discreteness of the momentum grid. For non-interacting particles, this results in an artificial finite size gap:
\begin{equation}
    \Delta_g[\bm{k}=(0,0)]_{\text{free}} = 8t_2\left[1 - \cos\left(\frac{2\pi}{L}\right)\right].
\end{equation}
For convenience, we assume a square lattice with $L_x = L_y = L$. This artificial contribution complicates the detection of true interaction-induced binding, especially in the regime of weak pairing, where $\Delta_g \ll \Delta_g[\bm{k}=(0,0)]_{\text{free}}$ or $L < \xi$.

To mitigate this problem we calculate the gap at shifted momentum $\bm{k} = (0, 2\pi/L)$, which eliminates the non-interacting gap. This momentum shift provides an additional benefit: it creates mixed boundary conditions (periodic along $x$, antiperiodic along $y$) for the relative wavefunction in Eq.~\ref{eq:fix momentum equation}. These boundary conditions generate bonding and antibonding states between two holes, with the antibonding state effectively making holes blind to the attraction and producing a continuum-like wavefunction (See Fig.~\ref{fig:momentum_correlation_heatmaps.pdf}(d)). The energy splitting between these states accurately reflects if the hole-hole interaction is attractive or repulsive.

\begin{figure}
\includegraphics[width=\columnwidth]{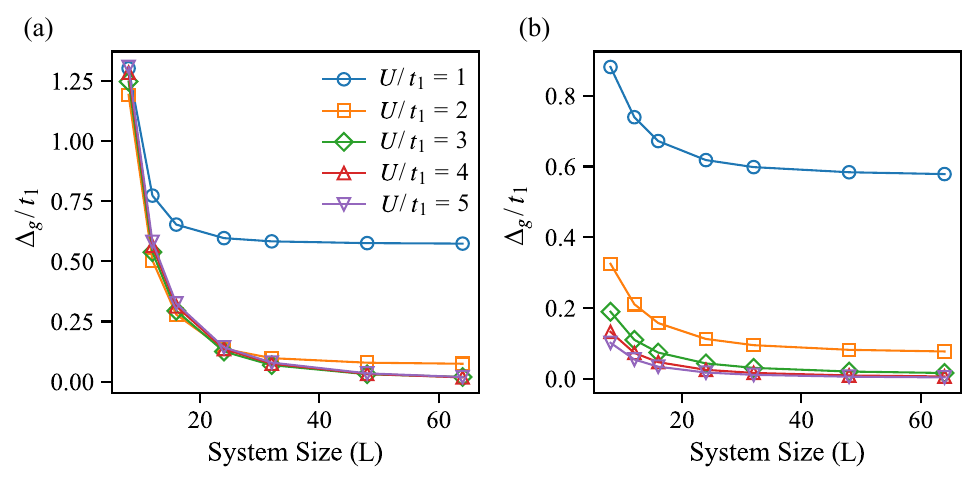}
\caption{\label{fig:gap_comparison_t2_-0.6.pdf} Comparison of binding energy calculations using different momentum sectors and system sizes. (a) Binding gap $\Delta_g(\bm{k}) = E_1(\bm{k}) - E_0(\bm{k})$ measured at $\bm{k} = (0, 0)$ for various system sizes and interaction strengths ($U/t_1 = 1-5$). (b) Numerically improved binding gap measured at $\bm{k} = (0, 2\pi/L)$, showing faster convergence with system size. }
\end{figure}

Fig.~\ref{fig:gap_comparison_t2_-0.6.pdf} demonstrates how our shifted-momentum approach improves convergence with system size. The shifted-momentum gap converges  faster than the conventional definition, allowing a more reliable extraction of pairing strength even from modest system sizes. We apply this approach consistently throughout our study of spin fluctuation effects on the binding gap.

\section{Hilbert space truncation}
\subsection{Motivation and methodology}

To capture the essential physics of antiferromagnetic polarons as pairing glue, we develop a targeted Hilbert space truncation strategy that selectively incorporates spin fluctuations in the vicinity of doped holes. This strategy enables exact diagonalization of systems large enough to study Cooper pair formation, while preserving physical accuracy for hole kinetics and computational tractability. In the spirit of the original method introduced by Trugman~\cite{trugman_interaction_1988,trugman_spectral_1990}, our method selectively incorporates quantum fluctuations in regions where they significantly affect hole pairing, while treating distant regions as classical Néel order in Section \ref{sec: lambda delta 00 limit}.

The physical rationale behind this approach is rooted in the principle of \emph{nearsightedness}, which holds that distant background spin fluctuations have a negligible effect on the interactions between holes. Specifically, spin fluctuations far from the holes contribute nearly identically regardless of the hole separation, and therefore cancel out in binding energy calculations. Moreover, the spin background retains strong N\'eel order due to the dominant $t_2$ hopping and antiferromagnetic exchange interactions.

By restricting quantum fluctuations to regions where they meaningfully influence hole dynamics—and treating more distant areas as a static N\'eel background—we systematically reduce the Hilbert space size while preserving the essential many-body correlations responsible for pairing. This approach eliminates irrelevant background contributions and captures the microscopic mechanisms underlying polaron binding. Similar truncation schemes have proven effective in other strongly correlated systems, such as electron-phonon coupled models~\cite{Nath2016}.

We implement this approach based on basic components of the QuSpin package~\cite{Weinberg2017a, weinberg_quspin_2019}, which provides efficient tools for constructing symmetry-conserving many-body Hamiltonians and facilitates controlled truncation of the Hilbert space.

\subsection{Single-hole basis construction}

Our Hilbert space construction begins with a systematic methodology for the single-hole case, which lays the foundation for addressing the complexities of the two-hole problem. The goal is to balance computational efficiency with physical accuracy through two essential steps.

First, we adopt the classical N\'eel-ordered state as the initial configuration, with spins aligned up on one sublattice and down on the other. This choice is motivated by our finding that kinetic effects alone can stabilize antiferromagnetic correlations, even in the absence of explicit spin exchange interactions. A single hole is then introduced into this ordered background.

Second, we expand the Hilbert space by incorporating controlled kinetic and spin-flip processes. We begin by allowing the hole to move through a single nearest-neighbor ($t_1$) or next-nearest-neighbor ($t_2$) hopping, thereby generating the minimal set of kinetically accessible configurations. These kinetic processes are recursively applied to all charge/spin configurations up to ${D_k}$ times. In our current approach, we set ${D_k} = 1$, whereas previous studies typically use ${D_k} \approx 3 \to 5$.  After these kinetic moves, we incorporate local spin fluctuations by applying spin-flip operators of the form $S_i^+ S_j^- + S_i^- S_j^+$, where $(i,j)$ are neighboring spins. These operators capture essential quantum magnetic dynamics in the vicinity of the hole. We restrict the spin-flip pairs to lie within the ``threshold distance'', an $L^\infty$ distance ${D_f}$ from the hole position. The use of the $L^\infty$ metric is motivated by the coexistence of NN and NNN hopping, and it is consistently adopted throughout our basis truncation scheme. This localization strategy captures the key physics of hole-induced spin correlations while avoiding exponential growth in the Hilbert space, resulting in a manageable single-hole basis of 370 states per momentum sector with ${D_f} = 2$.

\begin{figure}
\includegraphics[width=\columnwidth]{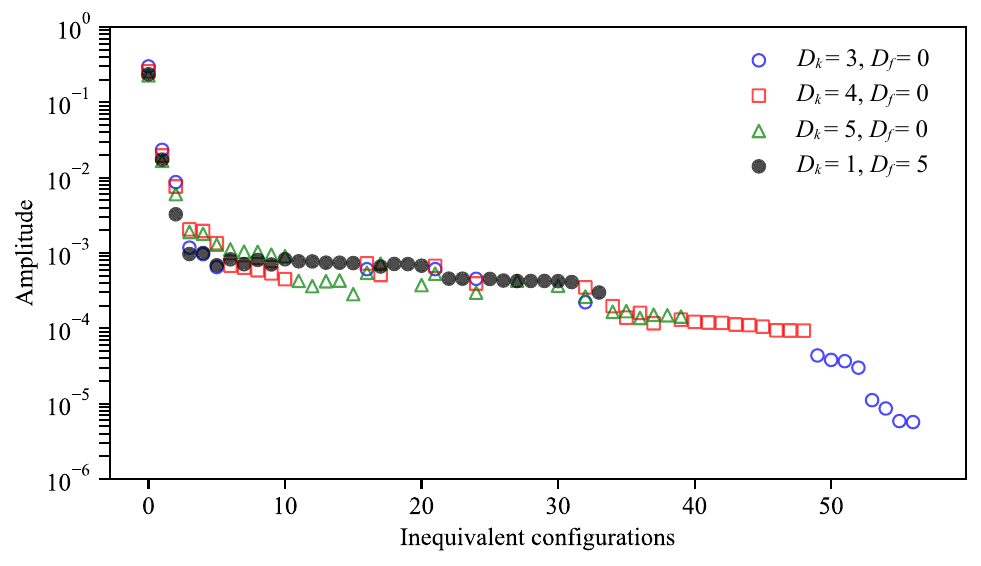}
\caption{\label{fig:single_hole_basis_spin_flip.pdf} Configurations with their amplitudes generated with various basis schemes. Only nonequivalent configurations (up to translations, rotations and reflections) with top 30 large amplitude are visualized for each scheme. The maximum distance between spin flips and holes for the improved method is set to be ${D_f} = 5$. We compare it against repeated kinetic propagation method with ${D_f} = 0$ , where the value ${D_k}$ indicates the number of kinetic propagation recursively applied in vertical, horizontal and diagonal directions to the classical N\'eel spin configurations. }
\end{figure}

Our approach represents a significant improvement over earlier recursive schemes based solely on repeated kinetic propagation~\cite{trugman_interaction_1988}. As illustrated in Fig.~\ref{fig:single_hole_basis_spin_flip.pdf}, our method---which combines one-step hole motion with targeted spin fluctuations---successfully captures many high-weight configurations that are inaccessible to recursive approaches. These include states with relatively simple spin textures but nontrivial hole trajectories, which contribute substantially to the ground state wavefunction and are missed when the basis is generated purely through kinetic processes.

\subsection{Two-hole basis construction}

\begin{figure*}
\includegraphics[width=1\linewidth]{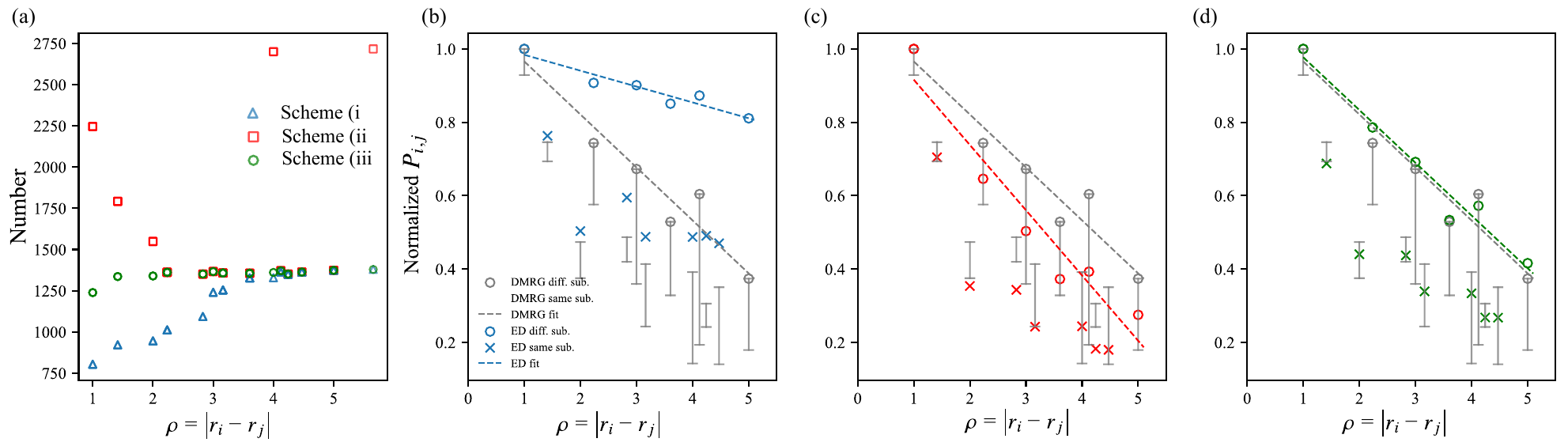}
\caption{\label{fig:combined_fluctuation_correlation.pdf} (a) Number of magnetic states per hole motion generated configuration of two holes. On one hand, fixing the range of magnetic fluctuations per individual hole tends to reduce the Hilbert when the distance between holes is smaller than the allowed range of fluctuations ${D_f}$ (because the same state is generated in more than one way), leading to effective repulsion. On the other hand, fixing the Hilbert space dimension  for arbitrary relative position (treating holes on the same/different sublattices separately) biases the system by generating an artificial  attractive potential. (b, c, d) Comparison between ED and DMRG for the hole-hole correlation function $P_{i, j} = \langle \bar{n}_{i} \bar{n}_{j} \rangle_G$ for $(\lambda, \Delta) = (1, 1)$ and $U = 10.0$, using i) the ``threshold distance'' ${D_f}$= 2, enlarging the Hilbert space dimension  ii) by fixing the Hilbert space dimension for any relative position of the two holes, and iii)  by making the Hilbert space dimension for each relative position between the two holes proportional to the number of magnetic states generated by hole motion (see main text). The third option gives the best match with the DMRG result. }
\end{figure*}

\begin{figure*}
\includegraphics[width=1\linewidth]{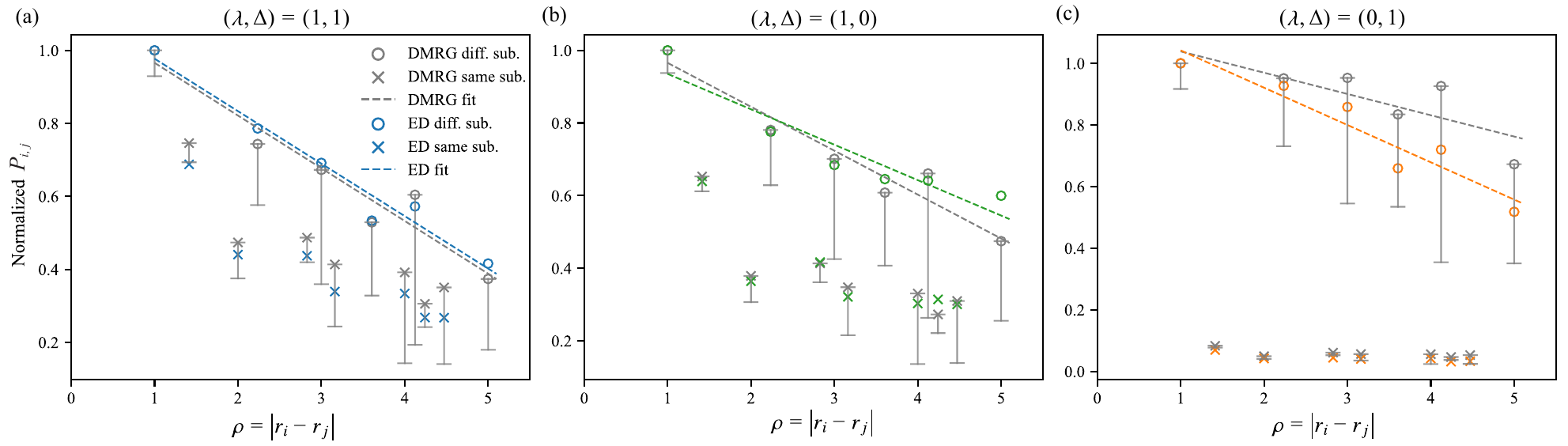}
\caption{\label{fig:ED_DMRG_three_parameters_comparison.pdf} (a) The ED/DMRG comparison in hole-hole correlation at $(\lambda, \Delta) = (1, 1)$ with $U=10t_1$. (b) Comparison at $(\lambda, \Delta) = (0, 1)$. (c) Comparison at $(\lambda, \Delta) = (1, 0)$. }
\end{figure*}

We now systematically extend the Hilbert space construction method developed for the single-hole case to the two-hole scenario, enabling unbiased energy comparisons across a wide range of hole configurations. While the construction follows similar guiding principles, additional care is required to control the Hilbert space dimension and avoid artificial biases in pairing energy estimates, as illustrated in Fig.~\ref{fig:combined_fluctuation_correlation.pdf}.

We begin by placing two holes into a classical N\'eel-ordered background, ensuring that they occupy opposite sublattices to maintain total $S_z = 0$. Each hole is then allowed to undergo a single hopping step, either nearest-neighbor ($t_1$) or next-nearest-neighbor ($t_2$), thereby generating the minimal set of kinetically accessible configurations. With the language of single hole case, this corresponds to ${D_k} = 1$ for both holes.

Next, we introduce local spin fluctuations around each hole. Each hole may be associated with at most one spin-flip pair, created by applying the bond operator $S_i^+ S_j^- + S_i^- S_j^+$ to the kinetically generated basis states, where $(i,j)$ are either nearest or second-nearest neighbors. We impose a maximum $L^\infty$ distance ${D_f}$ between the spin-flip pair and its associated hole to ensure that the resulting configurations can be decomposed into a product of two independent single-hole states when the holes are far apart. This constraint eliminates unphysical long-range correlations and ensures that the spin fluctuations are localized where they meaningfully impact hole dynamics.

A key challenge arises when comparing energies for different hole separations. Using a fixed maximum spin-flip distance typically yields a smaller Hilbert space dimension for adjacent holes compared to well-separated ones. This disparity introduces a systematic bias: configurations with larger Hilbert space of magnetic fluctuations benefit from greater variational freedom and thus artificially lower energies, obscuring the true physical nature of hole pairing.

This issue is compounded by the fact that our construction treats spin flips as independently associated with each hole, effectively forming a product basis. When the holes are close together, many configurations that appear distinct in this product space actually represent the same physical state due to overlap in the fluctuation regions. For example, spin flips assigned to one hole may equivalently be attributed to its neighbor. This redundancy reduces the effective Hilbert space for adjacent-hole configurations, artificially raising their energy and underestimating the pairing interaction. Physically, this manifests as an enforced classical ordering near the holes, which is energetically unfavorable relative to the true quantum ground state.

Fig.~\ref{fig:combined_fluctuation_correlation.pdf} quantifies this bias. In panel (a), we show the average fluctuation-generated Hilbert space per kinetic configuration, defined as the ratio of the fixed hole position subspace dimensions with and without spin fluctuations for fixed hole separations. When two holes are adjacent, the effective Hilbert space is nearly halved compared to the well-separated case due to configuration overlap. As a result, pairing tendencies are severely underestimated, as shown in panel (b).

To mitigate this bias, we implement a hierarchical prioritization scheme for generating spin-flip configurations. We begin by including all magnetic configurations generated with ${D_k} = 1$ and ${D_f} = 2$. As previously explained, when the two holes located at positions \( \bm{r}_i\) and \( \bm{r}_j\) are close, the ratio between the number of total magnetic configurations,  $N_{\text{fl} }(\rho)=N(D_k=1, D_f=2, \rho)$ with $\rho=|\bm{r}_i-\bm{r}_j|$ and the number of kinetically generated configurations,  $N_{\text{kin}}(\rho)=N(D_k=1, D_f=0, \rho)$, becomes significantly smaller. This motivates the increase of configurations due to magnetic fluctuations for small $\rho$ by  increasing \(D_f\). 

The extension is guided by a prioritization scheme that favors configurations in which the spin-flip pair is as close as possible to its associated hole. When multiple configurations share the same maximum hole spin-flip distance, we apply secondary and tertiary criteria based on the next-largest distances. All configurations meeting these criteria are included, and any configurations tied under the same ranking are  retained to preserve symmetry. We continue this process until adding further configurations would increase the maximal ratio $\max_{i, j} N_{\text{fl}}(\rho)/N_{\text{kin}}(\rho)$. This maximum ratio is fixed based on configurations with $D_k=1, D_f=2$, and we obtain the maximum separately for hole pairs on the same and different sublattices, acknowledging that sublattice effects remains relevant when the holes are far apart. This hierarchical approach favors compact configurations to address Hilbert space discrepancies, ensuring that the strongest hole-magnon interactions are fully captured.

We tested two basis enlargement schemes, one following the method illustrated above, and the other do in a much naive way, fixing $\max_{i, j} N_{\text{fl}}(\rho)$ with the same prioritization scheme.  Fig.~\ref{fig:combined_fluctuation_correlation.pdf} shows that our final scheme yields the best agreement with DMRG results. In contrast, simpler approaches, such as fixing ${D_k} = 1$ and ${D_f} = 2$ or fixing the Hilbert space dimension $N_{\text{fl}}(\rho)$ for any relative position of the two holes, lead to significantly poorer agreement. 

We note that the accuracy of our basis construction scheme is limited in certain regions of the $(\lambda, \Delta)$ parameter space. In particular, the approach is optimized for regimes where both $\lambda$ and $\Delta$ are simultaneously nonzero. When $\lambda = 0$ but $\Delta$ is large, the reliability decreases. In this limit, many kinetically generated states carry negligible amplitude due to the absence of $t_1$ hopping, which is essential for inter-sublattice hole motion. Nevertheless, for nearby hole configurations, it is still possible to generate kinetically generated states via spin fluctuations acting on the classical background—specifically, which gives identical configurations as both holes move to the opposite sublattice and create a single spin mismatch at their original positions. These processes occur exclusively when the holes are adjacent and can artificially enhance pairing in this regime, as shown in Fig.~\ref{fig:ED_DMRG_three_parameters_comparison.pdf}(c).

\subsection{Momentum sector  for superconducting gap }

\begin{figure}
\includegraphics[width=\columnwidth]{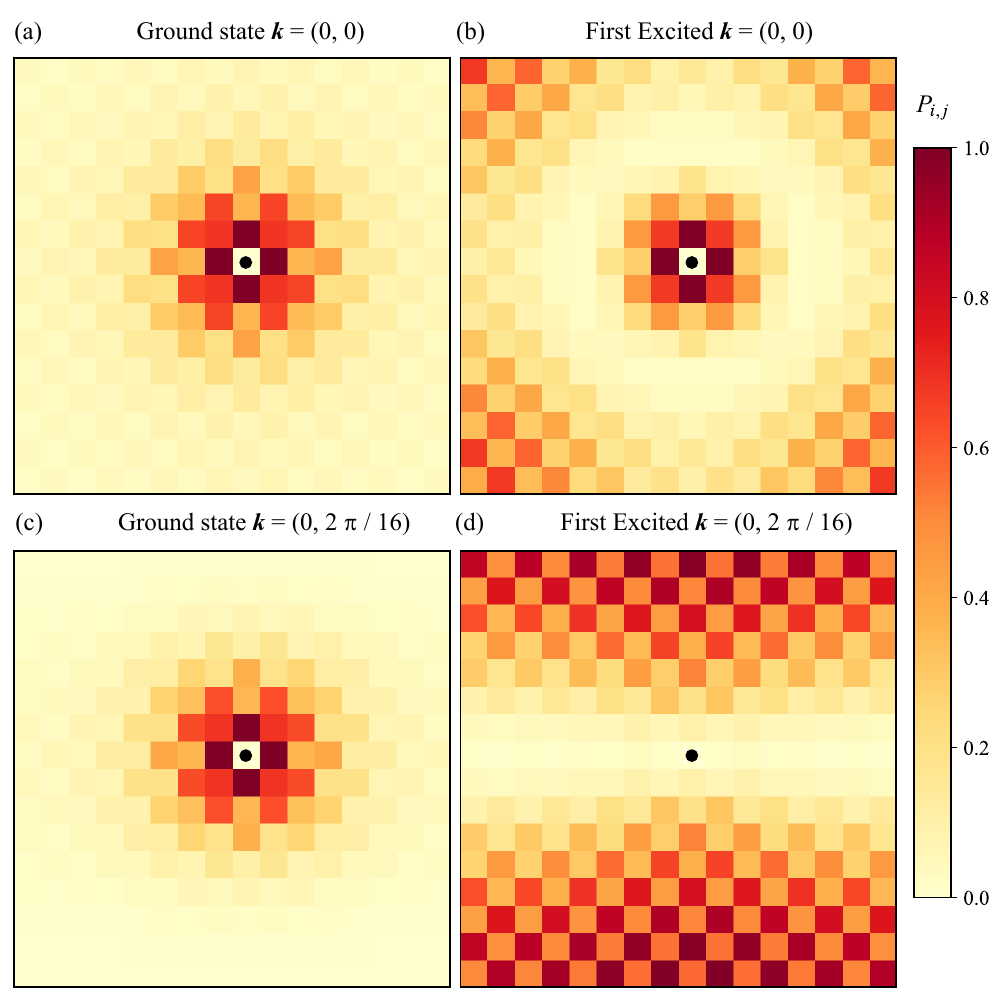}
\caption{\label{fig:momentum_correlation_heatmaps.pdf} Momentum sector dependence of hole-hole correlations demonstrates optimal conditions for superconducting gap calculations. We show correlation patterns $P_{i,j} = \langle\Psi_G | \bar{n}_{i} \bar{n}_{j} |\Psi_G\rangle$ in different energy eigenstates and momentum sectors, calculated using constrained ED with $(\lambda, \Delta) = (1, 1)$ and $U = 10t_1$. Position $i$ is fixed at the lattice center while $j$ varies across all sites. (a) Ground state at momentum $(0, 0)$ showing tightly bound Cooper pairs. (b) First excited state at momentum $(0, 0)$ exhibiting hybrid character with both binding and separated-hole components. (c) Ground state at momentum $(0, 2\pi/L)$ demonstrating preserved bound pair structure under mixed boundary conditions. (d) First excited state at momentum $(0, 2\pi/L)$ showing clean unbound state with minimal pairing correlations. The dramatic difference between excited states in different momentum sectors motivates our choice of $(0, 2\pi/L)$ for gap calculations. }
\end{figure}

We demonstrate that the careful selection of momentum sectors plays a crucial role in suppressing finite-size effects in superconducting gap calculations. By systematically comparing ground and excited states across different momentum sectors, we identify configurations that best preserve the physical nature of Cooper pairing while minimizing artifacts induced by boundary conditions.

The ground state exhibits tightly bound Cooper pairs across all momentum sectors, with characteristic coherence lengths extending over several lattice spacings. As shown in Fig.~\ref{fig:momentum_correlation_heatmaps.pdf}(a, c), the hole-hole correlation patterns remain largely unchanged when the momentum is varied from \((0,0)\) to \((0, 2\pi/L)\). This insensitivity highlights the robustness of pair formation and effectively mimics mixed boundary conditions---periodic in one direction and antiperiodic in the other---for the relative coordinate of the hole pair.

In contrast, the first excited state shows a pronounced sensitivity to the choice of momentum sector. At $\bm{k} = (0,0)$, it exhibits a hybrid character, combining signatures of bound-pair correlations with a significant probability amplitude for widely separated holes [Fig.~\ref{fig:momentum_correlation_heatmaps.pdf}(b)]. In the $\bm{k} = (0, 2\pi/L)$ sector, however, the excited state evolves into a nearly unbound configuration, effectively decoupled from the pairing potential [Fig.~\ref{fig:momentum_correlation_heatmaps.pdf}(d)]. This unbound state attains slightly lower energy than the hybrid configuration, providing a more accurate and less biased estimate of the superconducting gap.

\subsection{Pairing symmetry}

We identify the pairing symmetry of the superconducting ground state by systematically analyzing pair–pair correlation functions across different bond orientations. This approach allows us to distinguish between different pairing symmetries —such as $s$-wave and $d_{x^2 - y^2}$-wave— based on how pair correlations transform under lattice rotations and reflections.

Similar to the the formalism in Ref.~\cite{zhao_two-hole_2022}, we define the singlet pair operator as:
\begin{equation}
    \hat{\Delta}_{i j}^s=c_{i \uparrow} c_{j \downarrow}-c_{i \downarrow} c_{j \uparrow}
\end{equation}
We then calculate pair-pair correlation functions for parallel and perpendicular orientations:
\begin{eqnarray}
C_{i, j}^{s, \|} & =&\left\langle\Psi_G\left|\hat{\Delta}_{i, i+\hat{\bm{e}}_y}^s\left(\hat{\Delta}_{j, j+\hat{\bm{e}}_y}^s\right)^{\dagger}\right|\Psi_G\right\rangle, \\ 
C_{i, j}^{s, \perp} & =&\left\langle\Psi_G\left|\hat{\Delta}_{i, i+\hat{\bm{e}}_y}^s\left(\hat{\Delta}_{j, j+\hat{\bm{e}}_x}^s\right)^{\dagger}\right|\Psi_G\right\rangle    
\end{eqnarray}
Here we fix the reference pair $i, i+\hat{\bm{e}}_y$ at the lattice center and vary over all horizontal and vertical bonds that do not overlap with the central pair creation position.

\begin{figure}
\includegraphics[width=\columnwidth]{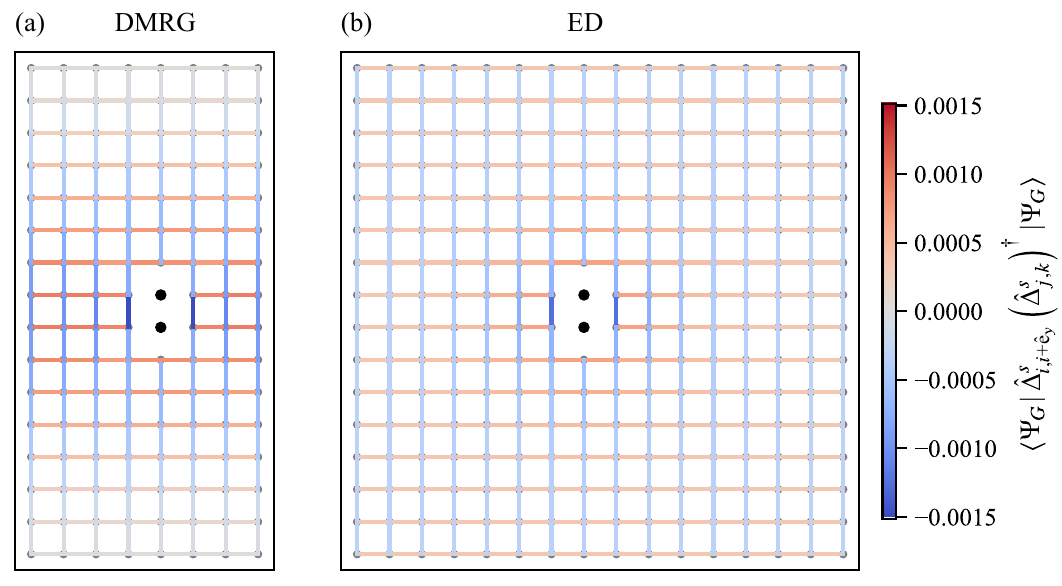}
\caption{\label{fig:combined_pair_pair_correlation.pdf} Pair-pair correlations confirm $d_{x^2-y^2}$ pairing symmetry through characteristic sign changes under rotation. We calculate $\left\langle\Psi_G\left|\hat{\Delta}_{i, i+\hat{\bm{e}}_y}^s\left(\hat{\Delta}_{j, k}^s\right)^{\dagger}\right|\Psi_G\right\rangle$ with reference pair fixed at lattice center ($i$) and test pairs at all nearest-neighbor bonds $(j, k)$. The negative relative sign  between horizontal pairs ($k = j + \hat{\bm{e}}_x$) and vertical pairs ($k = j + \hat{\bm{e}}_y$) confirms $d_{x^2-y^2}$ symmetry. (a) DMRG calculation on 8-leg cylinder with $(\lambda, \Delta) = (1, 1)$ and $U = 10t_1$. (b) Constrained ED calculation on $16\times16$ periodic cluster with identical parameters, scaled by factor of 4 to account for different pair densities. }
\end{figure}

Fig.~\ref{fig:combined_pair_pair_correlation.pdf} demonstrates clear $d_{x^2-y^2}$ symmetry signatures in both DMRG and ED calculations. Away from the pair annihilation position, the correlation amplitude changes sign upon rotating the pair creation operator by $\pi/2$. The amplitudes remain nearly identical in ED calculations, differing by only approximately 20\% in DMRG due to boundary conditions that break $C_4$ symmetry. This sign change upon $90^{\circ}$ rotation provides the characteristic signature of $d_{x^2-y^2}$ pairing symmetry.

\begin{figure}
\includegraphics[width=\columnwidth]{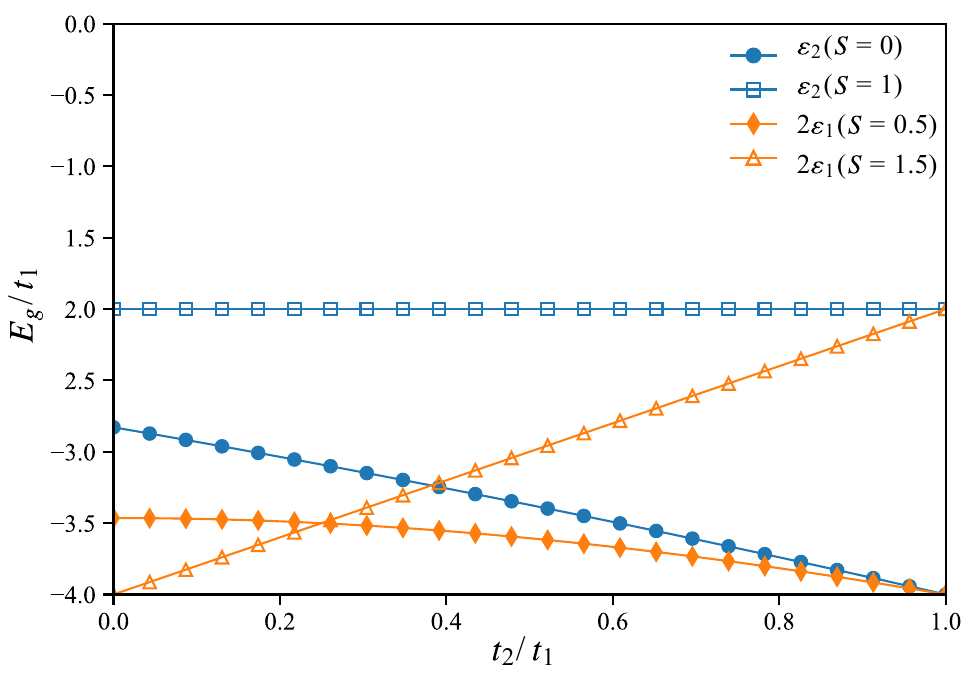}
\caption{\label{fig:detailed_binding_analysis.pdf} Ground state energies of the four-site model for different particle numbers and total spin quantum numbers $S$. Lower spin sectors consistently exhibit lower energies at $t_2/t_1 = 0.6$. Energy differences are defined as $\epsilon_1 (S) = E_{\text{1hole}} (S) - E_{\text{0holes}}$, where $E_{\text{0holes}} = \min_S E_{\text{0 hole}}(S)$,  and $\epsilon_2 (S) = E_{2 \; \text{holes}} (S) - E_{\text{0holes}}$. }
\end{figure}

\section{Singlet formation enhancement under hole doping}

Here we present the spin-spin correlation of the polaron with the doped hole fixed in position. Overall, as the ratio \(t_2/t_1\) increases from \(0\) to \(0.6\), singlet tendencies emerge and replace the original triplet correlations. Furthermore, when two holes form a bound pair, the singlet tendency is cooperatively enhanced. This behavior is consistently captured across exact diagonalization on small clusters, DMRG, and truncated Hilbert space ED methods.

A minimal four-site model captures the energy reduction in kinetic energy when forming spin singlet state. 
The ground states for different spin sectors are shown in Fig.~\ref{fig:detailed_binding_analysis.pdf}. While the fully polarized state has the lowest energy in the single-hole sector at $t_2/t_1 = 0$, the lowest-spin sectors consistently become energetically favored for $t_2/t_1 > 0.25$ in both the one- and two-hole cases, confirming that singlet formation is preferred. As expected, the stability of these low-spin states increases with $t_2/t_1$, supporting the interpretation that kinetic frustration promotes singlet formation near the holes.

\begin{figure}
\includegraphics[width=\columnwidth]{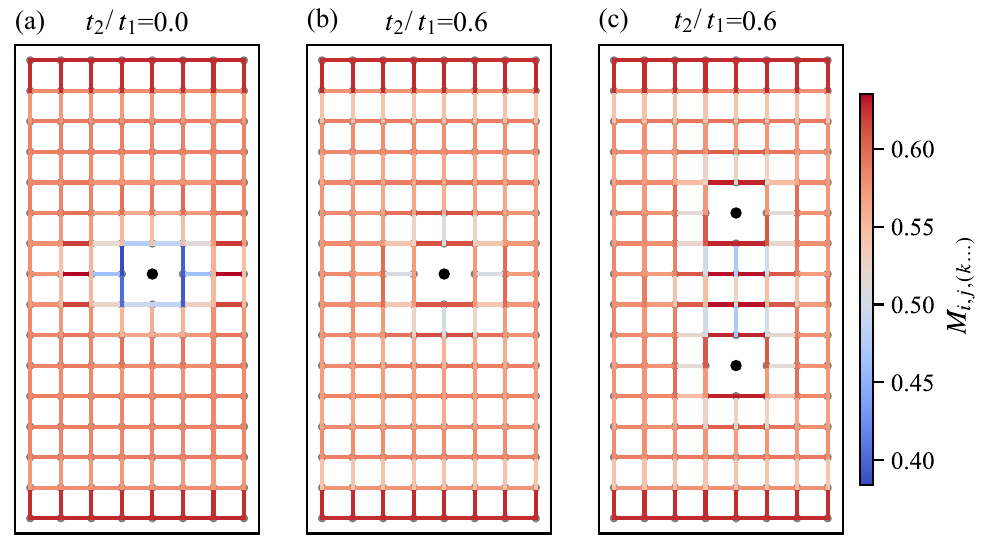}
\caption{\label{fig:dmrg_spin_correlation_comparison} We show correlations $M_{i,j,(k)}$ between sites $i$ and $j$ when a hole is fixed at position $k$ (black dot). (a) DMRG calculation on $8 \times 16$ cylinder with $(\lambda, \Delta) = (1, 1)$ and $U = 10t_1$ with $t_2 = 0.0$. (b) DMRG calculation on $8 \times 16$ cylinder with $(\lambda, \Delta) = (1, 1)$ and $U = 10t_1$ with $t_2 = 0.6$. (c) DMRG calculation for $M_{i,j,(k, l)}$ with same parameters as (b), but with two hole doped at $k, l$. }
\end{figure}

A similar trend is observed in larger lattices. As shown in Fig.~\ref{fig:dmrg_spin_correlation_comparison}, increasing $t_2/t_1$ from \(0\) to \(0.6\) suppresses the strong triplet correlations, initially characterized by $M_{i,j,(k)} \approx 0.37$, and instead stabilizes antiferromagnetic correlations with $M_{i,j,(k)} \approx 0.61$. This AFM correlation persists even in the two-hole case when the holes are doped but remain spatially separated. While there is a modest enhancement of singlet correlations around each hole, it remains limited compared to the contribution arising from spin interactions alone.

\begin{figure}
\includegraphics[width=\columnwidth]{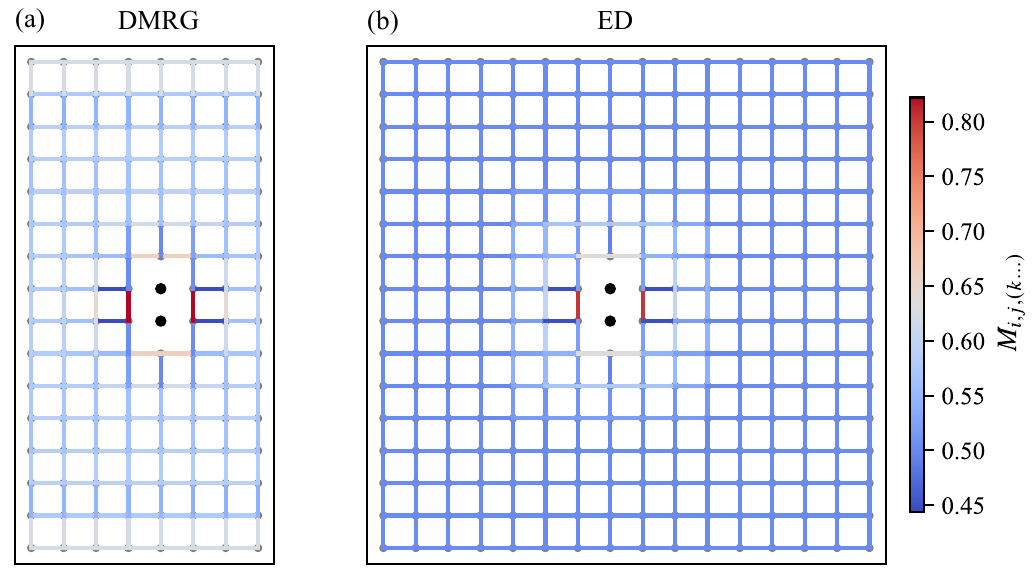}
\caption{\label{fig:combined_spin_correlation.pdf} We show correlations $M_{i,j,(k, l)}$ between sites $i$ and $j$ when a hole is fixed at position $k, l$ (black dot). (a) DMRG calculation on $8 \times 16$ cylinder with $(\lambda, \Delta) = (1, 1)$ and $U = 10t_1$. (b) Constrained ED calculation on $16\times16$ periodic cluster with identical parameters. }
\end{figure}

In contrast, when two holes occupy NN sites, singlet tendencies are significantly amplified—far exceeding the enhancement expected from spin-spin interactions alone (see Fig.\ref{fig:5}). Fig.~\ref{fig:combined_spin_correlation.pdf} illustrates this cooperative effect, showing an increase in singlet character from 0.58 to approximately 0.82, with excellent agreement between ED and DMRG results. The ED data match the DMRG findings closely, with a maximum deviation of only 3\% in the singlet correlation measured around the doped holes. 

The spin-spin correlation away from holes reveals the impact of spin fluctuation effects. As correlation measurements move away from holes, DMRG nearest-neighbor spin-spin correlations approach the pure Heisenberg model value of approximately 0.58, while ED results approach the classical Néel order value of 0.50 due to truncated basis states far from holes. However, ED calculations reveal enhanced spin-spin correlations at intermediate distances from holes, partially recovering DMRG fluctuation corrections through inclusion of local quantum fluctuations near the doped holes.

\section{Effective two-body hopping}

In this section, we demonstrate that singlet formation around fixed hole positions enhances the kinetic energy when two holes are nearby. In the main text, we map the kinetic energy contribution of the original model—which incorporates spin fluctuations within the full Hilbert space—onto a two-particle model with a two-body hopping term, i.e., a hopping operator that depends explicitly on the position of the other particle. Using the effective wavefunction $\psi(j, k)$, defined as the square root of the hole-hole correlation function $P_{j,k}$, this reformulation accurately reproduces the kinetic energy of the original full $t$-$J$ model.

The effective two-body hopping $|\tilde{t}_{jk} (l)|$ will never exceed $|t_{j, k}|$, owing to the Cauchy-Schwarz inequality: 
\begin{eqnarray} 
&&\langle \Psi_G | \bar{n}_{l} \bar{n}_{j} |\Psi_G\rangle \langle \Psi_G | \bar{n}_{l} \bar{n}_{k}
|\Psi_G\rangle
\geq |\langle \Psi_G|\bar{n}_{l} \sum_{\sigma} c_{k \sigma} c_{j\sigma}^\dagger \bar{n}_{l}|\Psi_G\rangle |^2. 
\nonumber \\
\end{eqnarray}

\begin{figure} \includegraphics[width=\columnwidth]{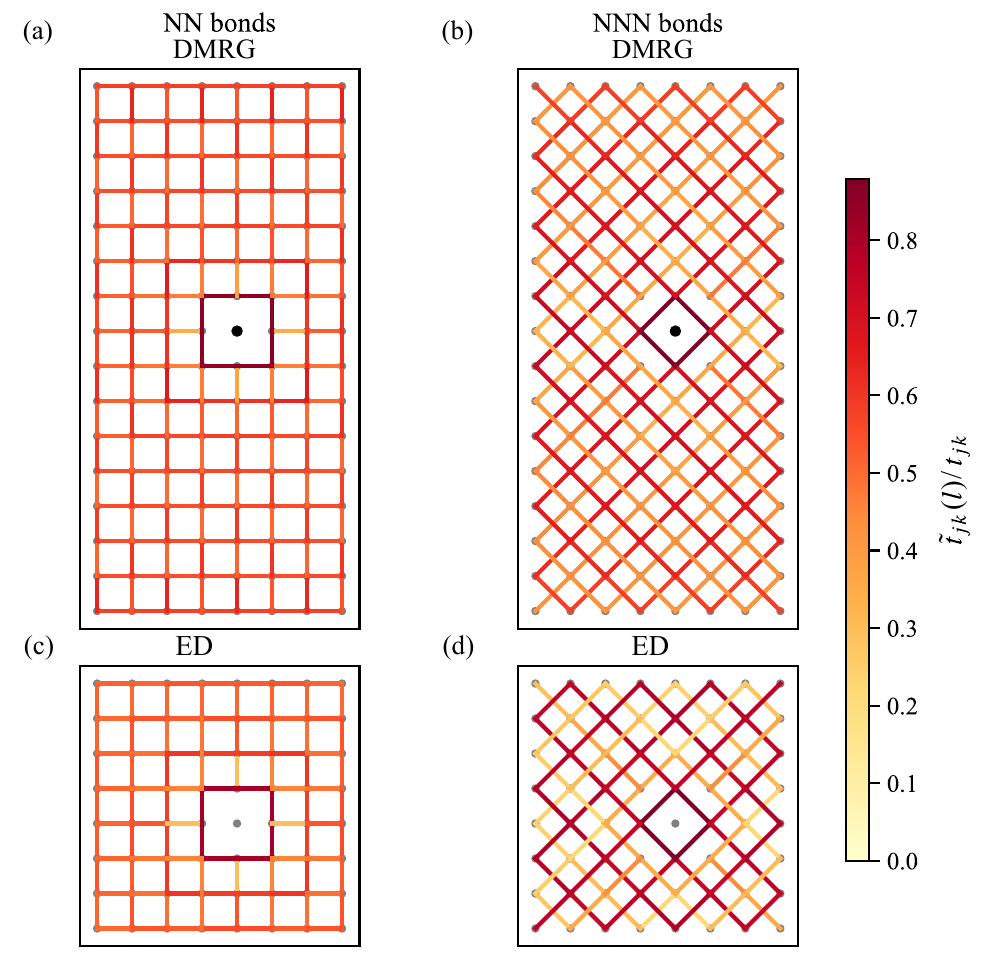} \caption{\label{fig:combined_hopping_visualization.pdf} DMRG-ED comparison for the two body hopping strength $\tilde{t}_{jk}(l)/t_{jk}$. DMRG calculations use the same parameters as in Fig.~\ref{fig:6} in the main text, and ED calculations are performed on an $8 \times 8$ cluster with $\lambda = \Delta = 1, U = 10 t_1$. (a, c) show vertical/horizontal hopping and (b, d) show diagonal hopping. (a, b) are DMRG results and (c, d) are ED results.} \end{figure}

A clear enhancement of both NN and NNN hopping amplitudes is observed in the vicinity of each hole. This effect is demonstrated using Hilbert space–truncated exact diagonalization on an $8 \times 8$ cluster with PBC, as well as DMRG calculations on an $8 \times 16$ cylindrical geometry, as shown in Fig.~\ref{fig:combined_hopping_visualization.pdf}.

\begin{figure}
\includegraphics[width=\columnwidth]{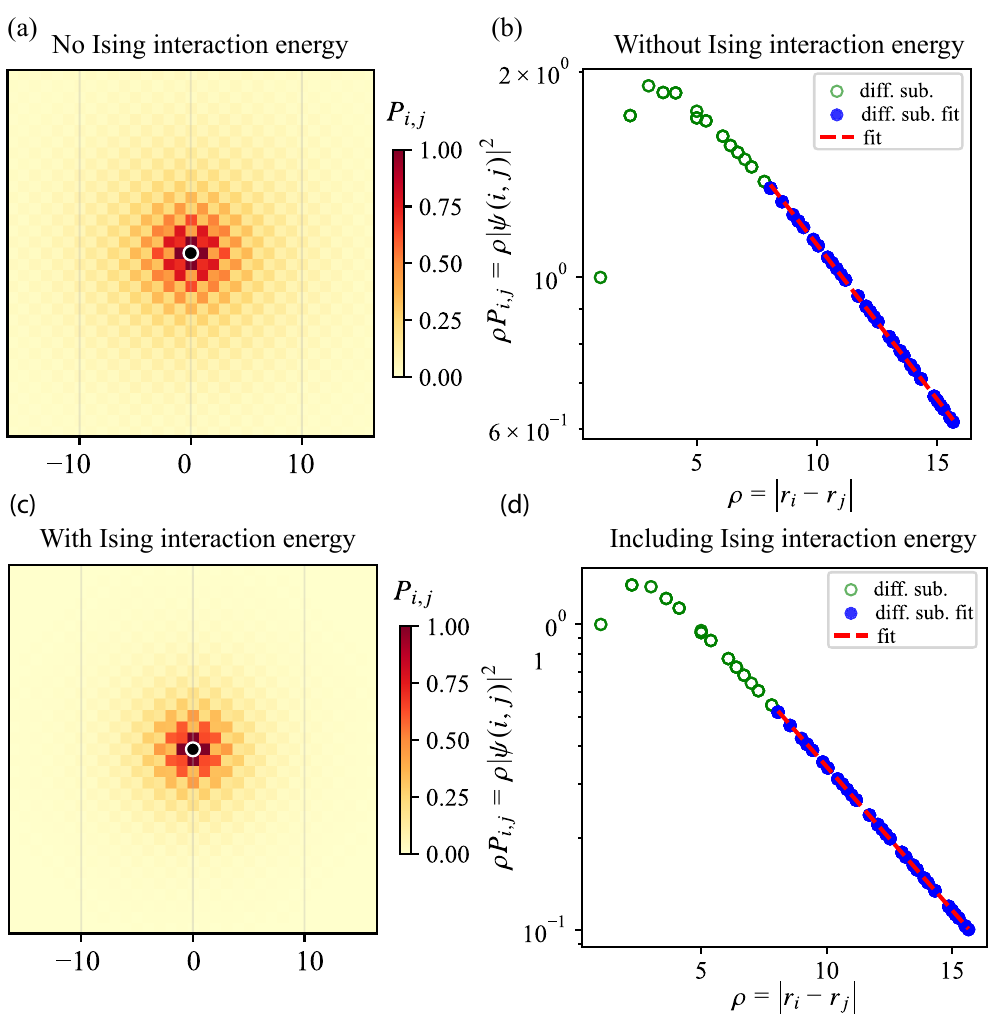}
\caption{\label{fig:variational_correlation_large.pdf}
(a) Hole-hole correlation \(P_{i,j} = |\psi(i, j)|^2\), calculated using two-body hopping amplitudes extracted from ED of the truncated Hilbert space on an \(8 \times 8\) system, without including the energy differences induced by Ising spin interactions. The calculation is performed with \(\lambda = \Delta = 1\) and \(U = 10t_1\). Only the central region is shown, though the full calculation spans a much larger region to approximate the thermodynamic limit. (b) Visualization of \(|r_{i} - r_{j}| P_{i,j}\) as a function of \(\rho = |r_{i} - r_{j}|\), based on the data in (a). (c) Hole-hole correlation \(P_{i,j}\) calculated with the same hopping enhancement as in (a), but including an attractive interaction of strength \(J_1/2\) from the Ising spin interaction. (d) Visualization of \(|r_{i} - r_{j}| P_{i,j}\) as a function of \({ \rho} \), based on the data shown in panel (c). }
\end{figure}

The enhancement of the effective two-body hopping amplitude at short hole-hole distances directly accounts for the formation of tightly bound Cooper pairs. We demonstrate this using the effective two-particle model extended to the thermodynamic limit, in which the two-body hopping amplitudes are extracted from our Hilbert space–truncated ED calculations, as shown in Fig.~\ref{fig:combined_hopping_visualization.pdf}(c, d). For large hole separations, we assign an average hopping value based on the bonds for the largest allowed hole-hole distance, as determined from the truncated ED data. Separate values are used for nearest-neighbor and next-nearest-neighbor hopping on different sublattices. The resulting hole-hole correlations from the two-particle model are shown in Fig.~\ref{fig:variational_correlation_large.pdf}(a). The coherence length $\xi$ is extracted by fitting the exponential decay of the Cooper pair wavefunction at large hole separation using the form:
\begin{equation}
    |\psi(i, j)|^2 \propto \frac{1}{\rho} e^{- \frac{\rho}{\xi}}
\end{equation}
where \(\rho\) is the distance between sites \(i\) and \(j\). As shown in Fig.~\ref{fig:variational_correlation_large.pdf} (b), Correlations are well captured by the fitting with coherence length \(\xi=9.6\)

To highlight the role of Ising spin interactions in pairing, we also present results where an additional attractive interaction of strength \(J_1/2\) between the two holes is included. These are shown in Fig.~\ref{fig:variational_correlation_large.pdf} (c, d) with coherence length \(\xi=4.6\).

\section{DMRG calculation details}
In this section, we provide detailed information on our U(1) DMRG calculations and compare our results across multiple interaction strengths and bond dimensions. 

We implement the ITensors library two-site DMRG algorithm on an 8-leg cylinder with $L_x=16$ \cite{fishman_itensor_2022, fishman_codebase_2022}. We impose periodic boundary conditions along the $L_y$ direction and open boundary conditions along the $L_x$ direction. Throughout the simulations, we conserve both the total electron number and the total spin projection $S_z$ quantum number, keeping up to 8000 states per block.

The presence of correlated hopping terms significantly increases the computational complexity compared to conventional $t$-$J$ models. These terms introduce approximately seven times more operators in the Hamiltonian and generate longer-range correlations, which in turn increase the cost of tensor contractions both in time and memory. To manage the computational complexity, we set a convergence threshold of energy differences below $10^{-6}$ between successive sweeps and progressively increase the bond dimension. Our calculations achieve truncation errors around $3 \times 10^{-5}$. Complete table of energies, bond dimensions and truncation errors for DMRG under different parameters are shown in table~\ref{tab:1}.

We select a longer length $L_x = 16$ relative to $L_y$ in order to minimize open boundary effects on hole density distribution. Open boundaries suppress hole density on the edges, reducing the likelihood of hole-hole correlations involving boundary-located holes. This boundary effect accounts for the observed breaking of $C_4$ symmetry in correlation patterns. 

\begin{figure}
\includegraphics[width=\columnwidth]{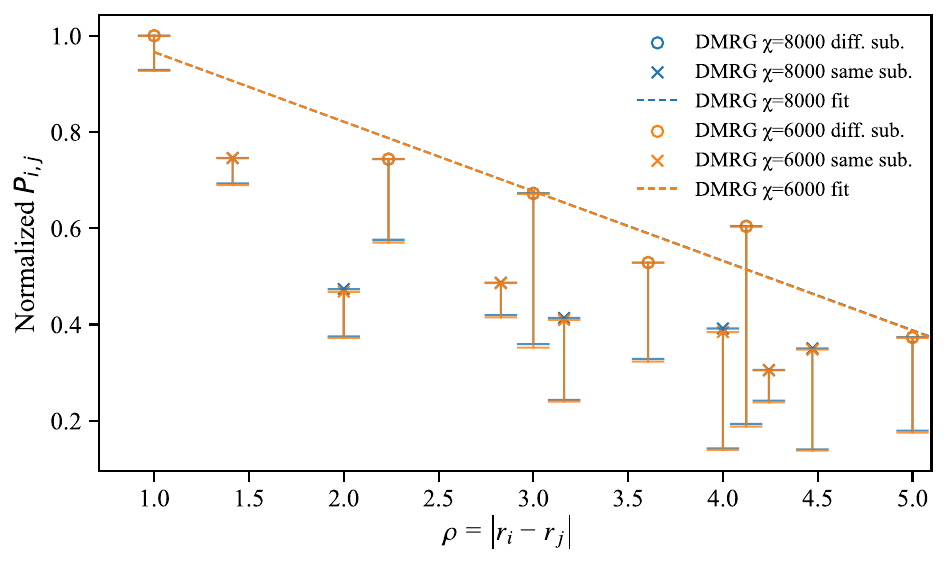}
\caption{\label{fig:dmrg_bond_dimension_comparison.pdf} DMRG bond dimension convergence validates numerical accuracy of hole-hole correlation calculations. We show correlations computed on the central $8\times 8$ region of an 8-leg cylinder with $(\lambda, \Delta) = (1, 1)$ and $U = 10.0 t_1$. Results obtained with bond dimensions $m = 6000$ and $m = 8000$ demonstrate excellent agreement, confirming numerical convergence. The slight overestimation at smaller bond dimension reflects entropic bias toward localized configurations.}
\end{figure}

Fig.~\ref{fig:dmrg_bond_dimension_comparison.pdf} demonstrates numerical convergence by comparing results obtained with bond dimensions 6000 and 8000. The two datasets show excellent agreement, with smaller bond dimensions tending to modestly overestimate the pairing strength. This mild and systematic bias originates from the entropic preference for more localized hole configurations in DMRG calculations.

\begin{figure}
\includegraphics[width=\columnwidth]{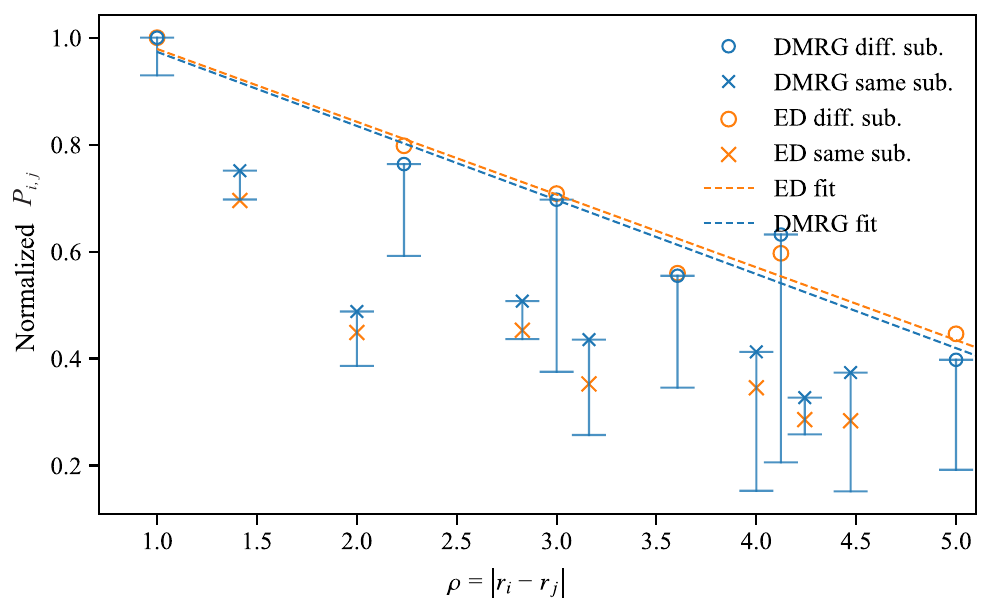}
\caption{\label{fig:correlation_vs_distance_U_12.0.pdf} The agreement between ED and DMRG  persists for higher interaction strength $U/t_1$, validating our computational approach. We compare the hole density-density correlation function versus distance for $(\lambda, \Delta) = (1, 1)$ and $U = 12.0 t_1$. Error bars in the DMRG data represent correlation spreads at identical distances due to boundary effects that reduce hole density away from cylinder centers.}
\end{figure}

Finally, we benchmark our DMRG results against exact diagonalization for larger interaction strength $U = 12.0 t_1$. As shown in Fig.~\ref{fig:correlation_vs_distance_U_12.0.pdf}, both methods exhibit consistent correlation behavior with no qualitative differences, confirming the robustness of our results as a function of increasing $U/t_1$.

\begin{table}[htbp]
    \centering
    \begin{tabular}{cccccccc}
        \hline\hline
        $N_\mathrm{hole}$ & $U/t_1$ & $t_2/t_1$ & $\lambda$ & $\Delta$ & $E/t_1$ & $m$ & maxerr \\
        \hline
        2 & 10.0 & 0.6 & 1.0 & 1.0 & $-67.8436$ & 8000 & $2.5\times10^{-5}$ \\
        2 & 12.0 & 0.6 & 1.0 & 1.0 & $-57.7240$ & 8000 & $2.5\times10^{-5}$ \\
        2 & 10.0 & 0.6 & 1.0 & 0.0 & $-54.8591$ & 2000 & $6.1\times10^{-8}$ \\
        2 & 10.0 & 0.6 & 0.0 & 1.0 & $-64.6172$ & 5000 & $2.5\times10^{-5}$ \\
        1 & 10.0 & 0.6 & 1.0 & 1.0 & $-64.2384$ & 5000 & $1.1\times10^{-5}$ \\
        1 & 10.0 & 0.0 & 1.0 & 1.0 & $-60.5464$ & 5000 & $1.1\times10^{-5}$ \\
        3 & 10.0 & 0.6 & 1.0 & 1.0 & $-71.2710$ & 5000 & $4.4\times10^{-5}$ \\
        4 & 10.0 & 0.6 & 1.0 & 1.0 & $-74.7879$ & 5000 & $4.9\times10^{-5}$ \\
        \hline\hline
    \end{tabular}
    \caption{\label{tab:1} Summary of model parameters and energies. \(E\) represents the ground state energy, \(m\) stands for DMRG max bond dimension, and `maxerr' represents the maximum truncation errors in singular value decomposition within the DMRG algorithm}
    \label{tab:parameters}
\end{table}

\end{document}